\documentclass[10pt]{article}
\usepackage[dvipdfmx]{graphicx}
\usepackage{amsmath}
\usepackage{amssymb}
\usepackage{ascmac}
\usepackage{comment}
\usepackage[usenames]{color}
\usepackage{bm}
\evensidemargin=\oddsidemargin
\def\ds{\displaystyle}
\begin{document}
%=====================================================================
%\rightline{\number\year/\number\month/\number\day}
\begin{center}
\Large
Numerical study on fast spectral evolution due to double resonance and applicability of generalized kinetic equation
%Numerical study on fast spectral evolution due to double resonance in a simplified two-layer fluid system
%Numerical study on the fast spectral evolution in a simplified two-layer fluid system
\end{center}
\begin{flushright}
Mitsuhiro Tanaka (Gifu University, JAPAN)
%Mitsuhiro Tanaka (Professor Emeritus, Gifu University)
\end{flushright}

\begin{abstract}
It is known that for a two-layer fluid system, the kinetic equation governing the evolution of the spectrum
of the wave field given by the standard wave turbulence theory (WTT) breaks down due to the existence of a
\lq\lq double resonance", and the spectrum can evolve on a time scale much faster than that predicted by
the standard WTT.
In this study, using a simplified model for a two-layer fluid system, the applicability of the generalized
kinetic equation (GKE) for such a situation is examined numerically.
It is shown that the GKE can reproduce the appearance of a sharp peak in the surface wave spectrum due to double resonance,
but it overestimates the growth of the peak, consequently failing to quantitatively describe the temporal
evolution of the spectrum correctly, particularly near the double resonance point.
\end{abstract}

%\tableofcontents

%=====================================================================
\section{Introduction}
%=====================================================================
Two-layer fluid systems, in which two fluids of different densities form stratified layers, 
are widely observed in natural environments. Typical examples include the estuarine region where 
freshwater flows over denser seawater, thermoclines in low-latitude shallow seas, etc.
% and seiches frequently observed in temperate lakes 
\cite{Osborne-Burch1980, Turner1973}.

In general, nonlinear wave-wave interactions play a crucial role in the evolution of the wave field,
particularly through resonant interactions \cite{Phillips1960}.
For single-layer fluids, the dispersion relation of surface gravity
waves, $\omega = \sqrt{gk \tanh kh}$, does not support three-wave resonance.
Hence, resonant energy transfer can only occur via higher-order (four-wave) interactions,
which leads to weak energy exchange between different wavenumbers and correspondingly slow
spectral evolution \cite{Phillips1960}.

On the other hand, in two-layer fluid systems, various types of three-wave resonant interactions become possible,
allowing for stronger energy transfer \cite{Choi2021}. 
%Under the assumption of one-dimensional propagation, a specific type of three-wave resonance involving
%two surface waves and one interfacial wave, all propagating in the same direction, is classified
%as \lq\lq Class 3." Notably, this is the only configuration in which all three resonant waves propagate
%in the same direction when the density ratio $\rho_1/\rho_2>1/3$.
%
In our previous study \cite{Tanaka-Wakayama}, we investigated spectral energy transfer from surface to
interfacial waves using direct numerical simulations (DNS).
There, we observed a strange phenomenon in which a sharp peak appeared in the surface wave spectrum 
at a certain wave number and grows rapidly on a timescale much faster than that predicted by
the standard WTT, i.e., O($1/\omega_0 \epsilon^2)$,
where $\omega_0$ is the typical frequency and $\epsilon$ is a nondimensional parameter representing the smallness
of the amplitudes of waves.

%------------------------------------------------------------------
\begin{figure}[h]
    \begin{center}
    \includegraphics[width=0.4\linewidth]
    {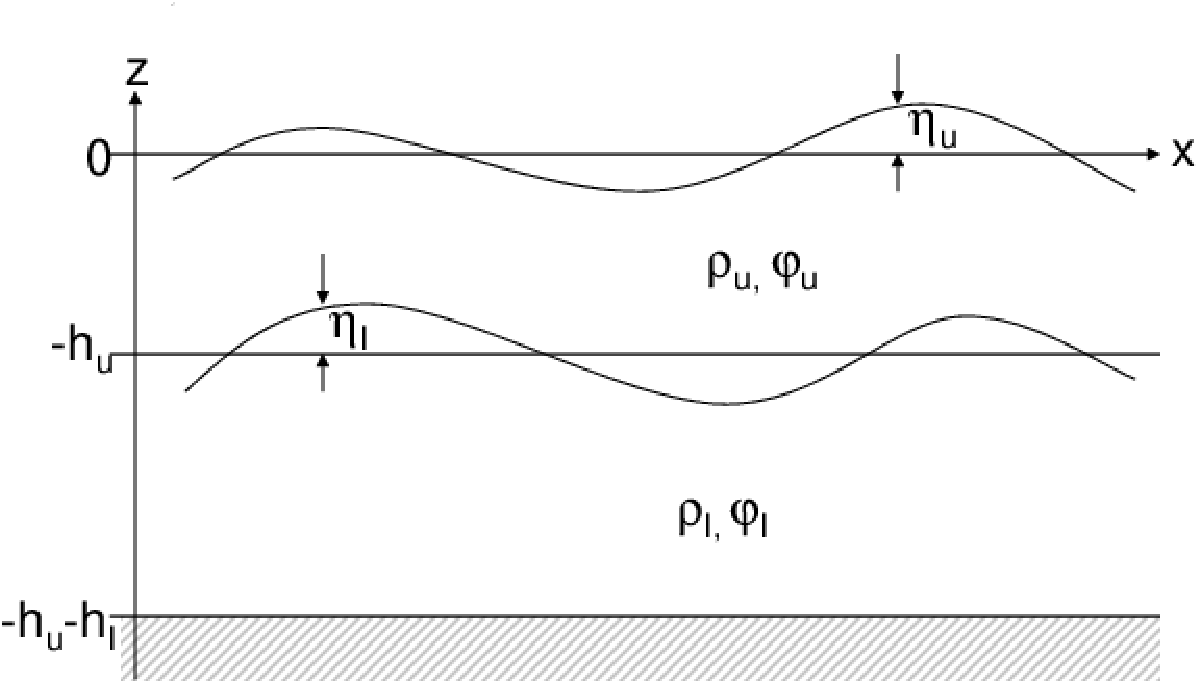}
    \end{center}
\caption{Schematic diagram of the two-layer fluid system}
\label{fig:definition}
\end{figure}
%------------------------------------------------------------------

A few years after the publication of the paper\cite{Tanaka-Wakayama}, we realized that this fast spectral change,
which was inconsistent with the standard WTT, was caused by \lq\lq double resonance".
On the other hand, relatively recently, various types of generalized kinetic equations (GKE) have been derived
that have greatly expanded the scope of applicability of the conventional kinetic equations.
These GKE do not assume slow spectral evolution and has been applied to describe rapid spectral changes
such as the one observed in the ocean wind waves caused by squall \cite{Annenkov_Shrira2015}.
Therefore, GKE may be able to describe the growth of the sharp peak due to double resonance,
which occurs on a faster time scale than that predicted by the standard WTT as observed in our previous work.
Examining this possibility is the aim of the present study.

The scenario of this paper is as follows.
In the rest of this introduction section, we review the results of our previous paper \cite{Tanaka-Wakayama}
which gave us the motivation of this paper.  There we also discuss the contradictions between the
results we obtained and the prediction based on the standard WTT, and also refer to the phenomenon of double resonance.
In \S2, we propose a \lq\lq simplified" model based on the Hamiltonian formalism of
the original two-layer fluid system in order to facilitate the numerical computations,
and derive the GKE for this simplified model.
We compare in \S3 the results of DNS and GKE for the simplified two-layered fluid system,
and evaluate the applicability of the GKE to the fast spectral changes induced by double resonance.
Summary and conclusions of the study are given in \S4.

%=====================================================================
\subsection{Review of our previous work}
%=====================================================================
In our previous work\cite{Tanaka-Wakayama}, we considered a two-layer fluid system as shown in Fig.~\ref{fig:definition}. 
Let the densities of the upper and lower layers be $\rho_u$ and $\rho_l$, respectively,
and their undisturbed depths are $h_u$ and $h_l$, respectively. 
The surface and interface displacements due to wave motion are denoted by $\eta_u(x,t)$ and $\eta_l(x,t)$, respectively. 
We set the $x$-axis in the direction of wave propagation and the $z$-axis vertically upward from
the undisturbed free surface.
The fluids are assumed to be incompressible and inviscid, and the flow is assumed to be irrotational, 
allowing the use of velocity potentials $\phi_u(x,z,t)$ and $\phi_l(x,z,t)$ for the upper and lower layers, respectively.
It is also assumed that the fluid motion is confined to a vertical plane, hence the wave propagation is restricted
to be  one-dimensional.

Then the governing equations for this system are given as follows:
\begin{subequations}
\begin{align}
& \nabla^2 \phi_u =0,\quad -h_u+\eta_l<z<\eta_u, \label{eqn:equations1}\\
& \nabla^2 \phi_l =0,\quad -h_u-h_l<z<-h_u+\eta_l, \\
& \eta_{u,t}+\eta_{u,x}\phi_{u,x}-\phi_{u,z} =0, \quad z=\eta_u, \\
& \phi_{u,t}+\frac{1}{2}\left(\phi_{u,x}^2+\phi_{u,z}^2\right)+g\eta_u =0, \quad z=\eta_u, \\
& \eta_{l,t}+\eta_{l,x}\phi_{u,x}-\phi_{u,z} =0, \quad z=-h_u+\eta_l, \\
& \eta_{l,t}+\eta_{l,x}\phi_{l,x}-\phi_{l,z} =0, \quad z=-h_u+\eta_l, \\
& \rho_u\left[\phi_{u,t}+\frac{1}{2}\left(\phi_{u,x}^2+\phi_{u,z}^2\right)+g\eta_l\right] \nonumber \\
&\quad
-\rho_l\left[\phi_{l,t}+\frac{1}{2}\left(\phi_{l,x}^2+\phi_{l,z}^2\right)+g\eta_l\right]=0,
\quad z=-h_u+\eta_l, \\
& \phi_{l,z} =0, \quad z=-h_u-h_l.
\label{eqn:equations8}
\end{align}
\label{eqn:governing equations}
\end{subequations}

By linearizing the governing system (\ref{eqn:governing equations}) of equations,
the following linear dispersion relation is obtained:
\begin{equation}
(1+R\,T_u\,T_l)\,\omega^4  - gk(T_u+T_l)\,\omega^2 +(1-R)g^2 k^2 \, T_u T_l=0,
\end{equation}
where
$\ds{T_u = \tanh kh_u,  \quad T_l = \tanh kh_l, \quad  R = \rho_u/\rho_l}$.
This equation is quartic in $\omega$, yielding four roots for each $k$: $\pm\omega^{+}$, $\pm\omega^{-}$,
where $\omega^{+} > \omega^{-}$.
\footnote{
Although the present study is restricted to one-dimensional propagation and thus $k$ is a scalar, 
it still represents a wavenumber vector in the broader physical context and takes values
in $(-\infty, \infty)$. On the other hand, the frequency $\omega$ is always taken to be positive; 
hence, the sign of $k$ indicates the propagation direction, and waves with $k$ and $-k$ are treated
as distinct and independent.
}

\begin{figure}[h]
    \begin{center}
    \includegraphics[width=0.4\linewidth]
    {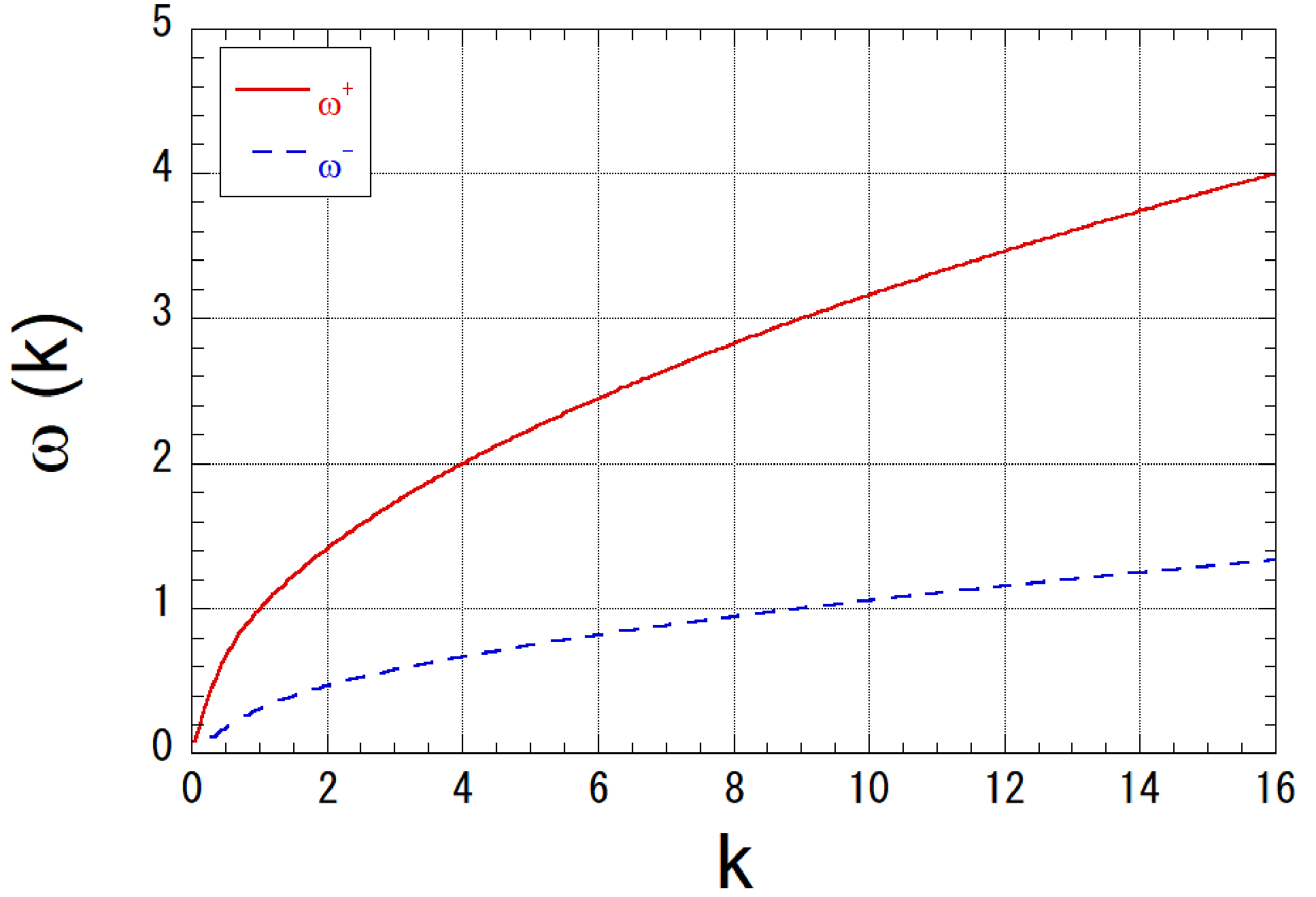}
    \end{center}
\caption{The linear dispersion relation ($h_l=2.0$, $R=0.8$)}
\label{fig:dispersion relation}
\end{figure}

The mode corresponding to the higher frequency $\omega^{+}$ is referred to as the \lq\lq surface wave mode",
as the surface displacement $\eta_u$ dominates over the interfacial displacement $\eta_l$.
Conversely, the lower frequency $\omega^{-}$ corresponds to the \lq\lq interfacial wave mode",
where $\eta_l$ dominates over $\eta_u$.
In this study, physical quantities are nondimensionalized such that $\rho_l = 1$, $h_u = 1$, and $g = 1$.
Therefore, the system is identified by two nondimensional parameters:
the density ratio $R \,(\equiv \rho_u/\rho_l<1)$ and the depth ratio $h_l\, ( = h_l/h_u )$.
As an example, we show in Fig.\ref{fig:dispersion relation} $\omega^{+}$ and $\omega^{-}$ when $h_l=2.0$ and $R=0.8$.
%------------------------------------------------------------------
%------------------------------------------------------------------

In our previous paper\cite{Tanaka-Wakayama}, the initial wave field was constructed by superposing huge number of
 monochromatic wavetrains with different wavenumbers so that the energy spectra of the surface wave mode $S^{+}(k)$
and the interfacial wave mode $S^{-}(k)$ were given by 
%---------------------------------------------------------------------
\begin{subequations}
\begin{align}
& S^{+}(k)=A\left(\frac{k}{k_p}\right)^{-3}\exp\left[-\frac{5}{4}\left(\frac{k}{k_p}\right)^{-2}\right]
\quad (k>0), \qquad S^{+}(k)=0 \quad (k<0), \\
\noalign{\bigskip}
& S^{-}(k)=0 \quad (-\infty<k<\infty).
\end{align}
\label{eq:initial spectrum}
\end{subequations}
%---------------------------------------------------------------------
As shown in (\ref{eq:initial spectrum}), the initial wave field consisted solely of a superposition
of surface wave modes propagating in the positive $x$-direction, and there were no interfacial wave modes at all.
The form of $S^{+}(k)$ was borrowed from the Pierson-Moskowitz spectrum, a well-known spectrum
for wind waves, but this spectral shape had no particular significance for the study.

\begin{figure}
\begin{minipage}{0.5\linewidth}
\begin{center}
    \includegraphics[width=0.9\linewidth]
    {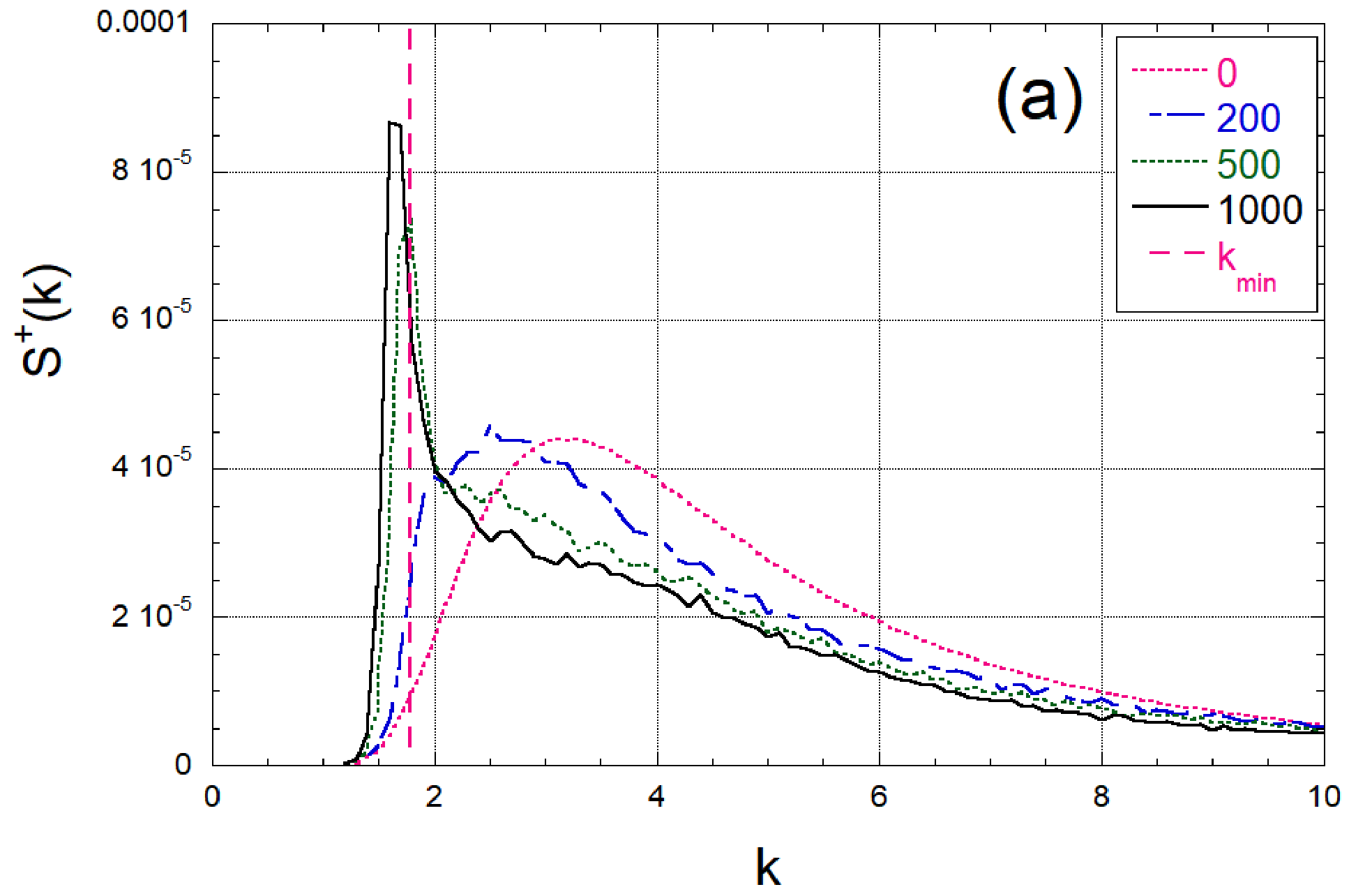}
\end{center}
\end{minipage}
\begin{minipage}{0.5\linewidth}
\begin{center}
    \includegraphics[width=0.9\linewidth]
    {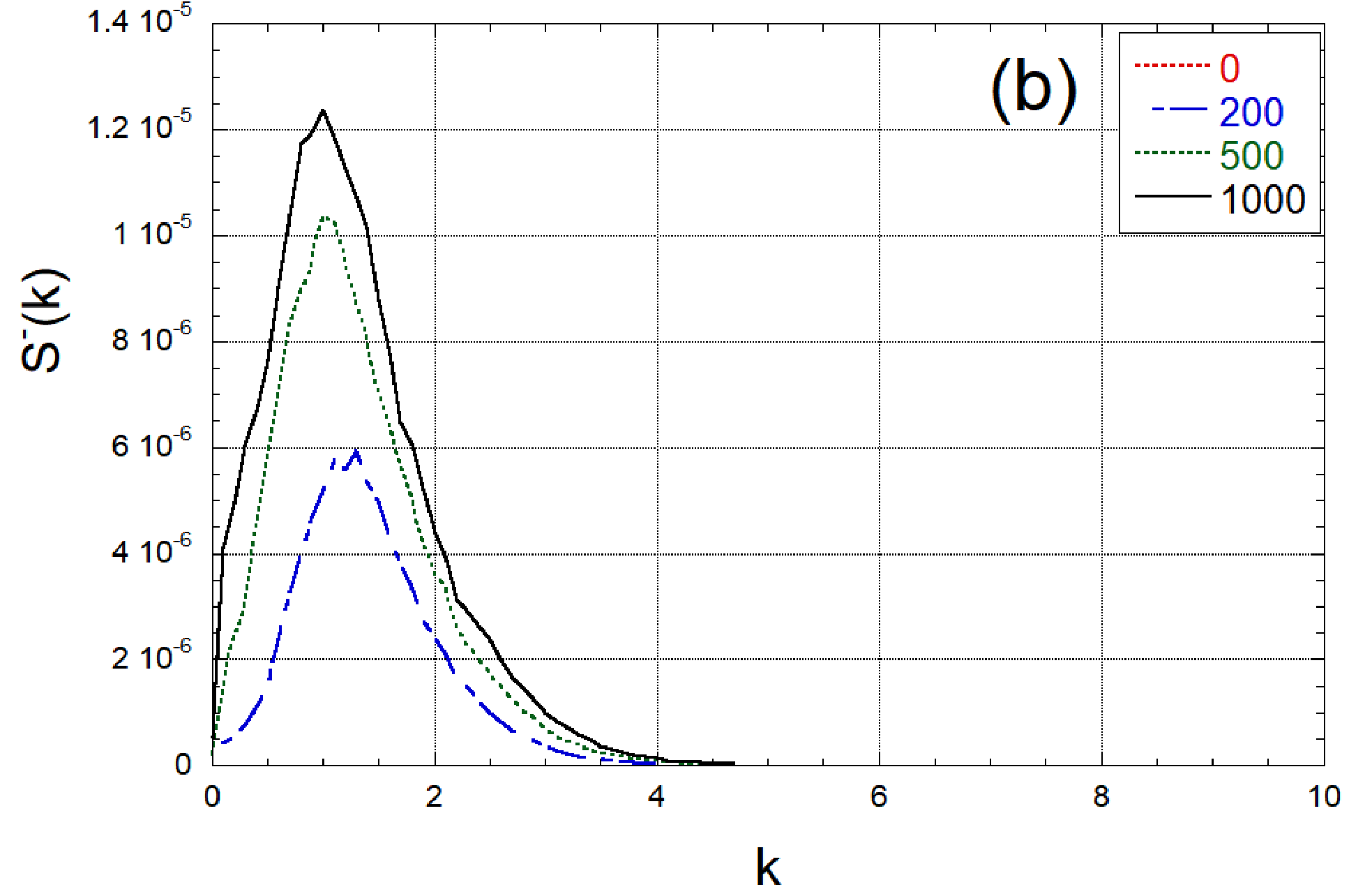}
\end{center}
\end{minipage}
\caption{Time evolution of the energy spectra when $h_l=2.0$, $R=0.8$, $k_p=3.5$.
$t=0$, $200T_p$, $500T_p$, $1000T_p$.
(a) Surface wave mode, (b) Interfacial wave mode.}
\label{fig:Spectra_R=0.8}
\end{figure}

Figure \ref{fig:Spectra_R=0.8} shows the time evolution of the spectra $S^{+}(k)$ and $S^{-}(k)$ obtained by DNS
of the governing equations (\ref{eqn:governing equations}) starting from a wave field with the initial
spectrum (\ref{eq:initial spectrum}).
In the figure, $T_p$ denotes the period corresponding to the surface wave mode with the peak wavenumber 
$k_p$ of the initial spectrum.  
The parameter $A$, determining the energy density of the initial wave field, 
was set to $A=1.5\times 10^{-4}$ (equivalent to $ak\approx 0.08$).  
The numerical method of DNS was based on the Higher-Order Spectral Method (HOSM) developed
by Alam et al.(2009)\cite{Alam_etal2009}.
For more details of the numerical method, including the method of estimating the spectra from those variables
that are directly treated by the HOSM, see \cite{Tanaka-Wakayama}.
%------------------------------------------------------------------
%------------------------------------------------------------------

As seen in Fig.~\ref{fig:Spectra_R=0.8}, the surface wave spectrum $S^{+}(k)$ gradually downshifted toward lower
wavenumbers during the initial stage of spectral evolution.
However, this downshift stopped at some time, and a sharp peak emerged around $k=1.73$ and grew rapidly.
On the other hand, the interfacial wave spectrum $S^{-}(k)$, which was initially zero at all wavenumbers,
monotonically grew in small wavenumber region around $k\approx 1$.
%------------------------------------------------------------------
\begin{figure}
\begin{minipage}{0.5\linewidth}
\begin{center}
    \includegraphics[width=0.9\linewidth]
    {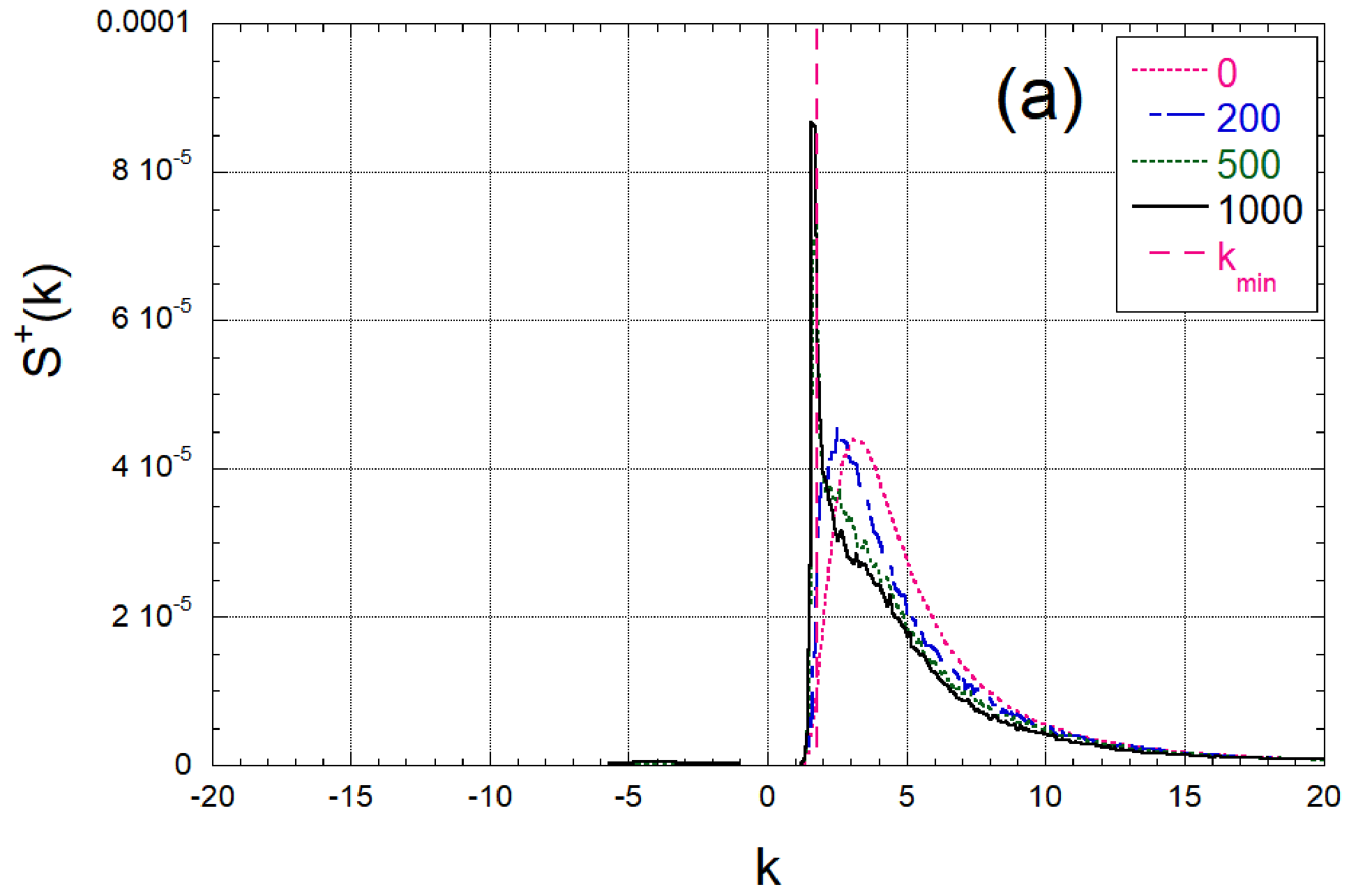}
\end{center}
\end{minipage}
\begin{minipage}{0.5\linewidth}
\begin{center}
    \includegraphics[width=0.9\linewidth]
    {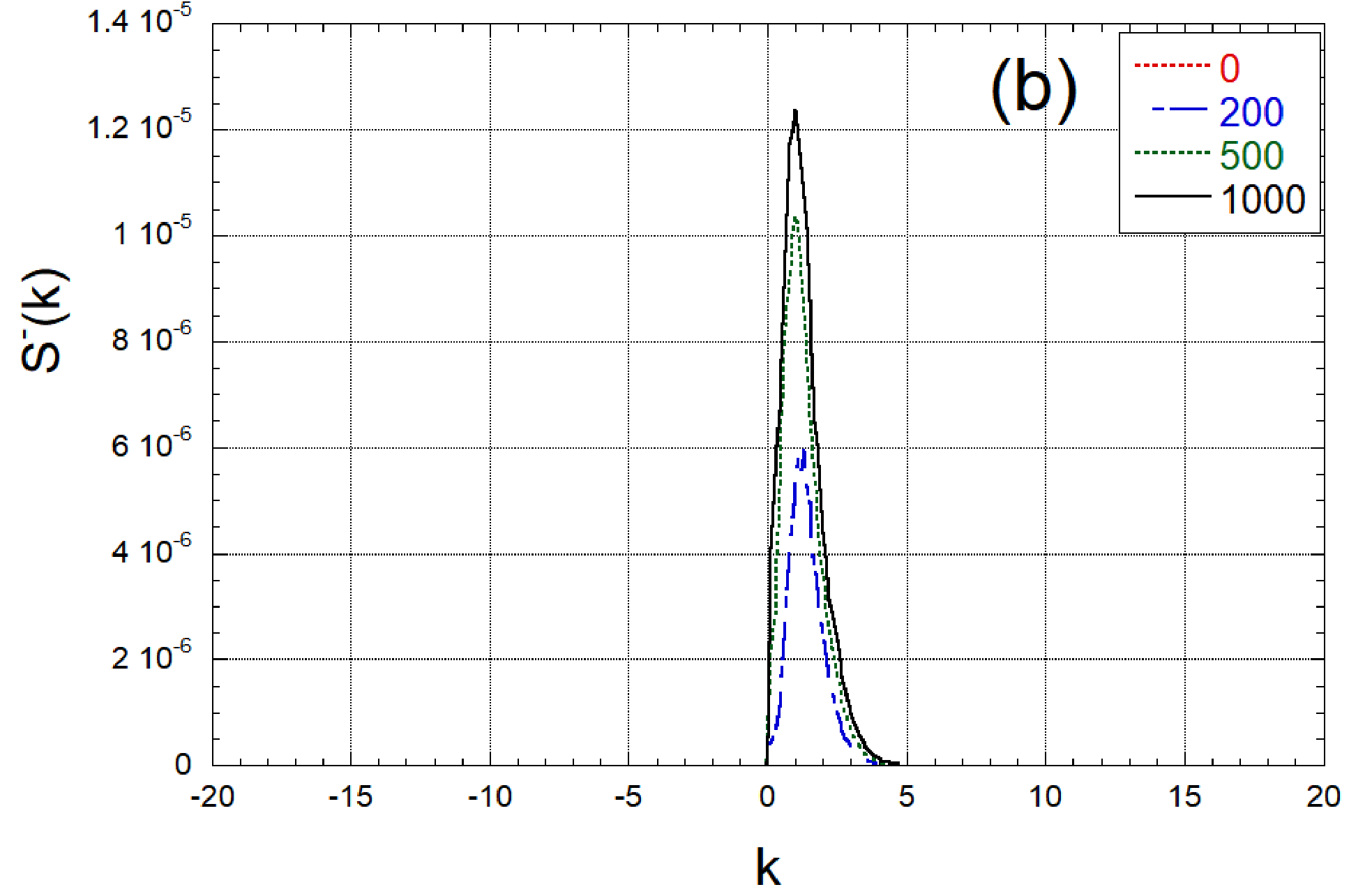}
\end{center}
\end{minipage}
\caption{\small Same as Fig.\ref{fig:Spectra_R=0.8} but for a wider range of $k$.
Note that waves propagating in the negative direction has been scarcely excited.}
\label{fig:Spectra_R=0.8_wide}
\end{figure}
%------------------------------------------------------------------
Figure \ref{fig:Spectra_R=0.8_wide} shows the identical $S^{+}(k)$ and $S^{-}(k)$ as shown in 
Fig.~\ref{fig:Spectra_R=0.8} over a wider range of $k$,
clearly demonstrating that both $S^{+}(k)$ and $S^{-}(k)$ remained virtually zero for $k<0$ up to $t=1000T_p$.
This indicates that waves propagating in the negative direction were scarcely excited,
and the wave field still consisted only of waves propagating in the positive $x$-direction even after 1000 periods have past.
%=====================================================================
\subsection{Contradiction with conventional WTT}
%=====================================================================
Applying the standard procedures of WTT (see, for example, \cite{Nazarenko2011,Zakharov et al.2025}),
such as assuming weak non-Gaussianity and $t\gg 1/\omega_0$, to a \lq\lq simplified" two-layer fluid system 
(\ref{eqn:simple model}) which will be introduced below yields the following kinetic equations for the spectra,
\begin{subequations}
\label{eqn:kinetic equations}
\begin{align}
\frac{dn^{+}_k}{dt}&=-2\pi\iint \bigg[ 
|V^{(2)}_{k12}|^2 \{(n^{+}_k-n^{+}_1)n^{-}_2 +n^{+}_k n^{+}_1\} 
\delta(\omega^{+}_k-\omega^{+}_1-\omega^{-}_2)\, \delta(k-k_1-k_2) \nonumber\\
&-|V^{(2)}_{1k2}|^2 \{(n^{+}_1-n^{+}_k)n^{-}_2 +n^{+}_k n^{+}_1\}
\delta(\omega^{+}_k-\omega^{+}_1+\omega^{-}_2)\, \delta(k-k_1+k_2)\bigg] dk_{12},  \\
\noalign{\bigskip}
\frac{dn^{-}_k}{dt}
&=2\pi\iint
|V^{(2)}_{12k}|^2 \{(n^{+}_1-n^{+}_2) n^{-}_k +n^{+}_1 n^{+}_2\}  
\delta(\omega^{-}_k-\omega^{+}_1+\omega^{+}_2)\, \delta(k-k_1+k_2) dk_{12},
\end{align}
\end{subequations}
where $n^{+}(k)$ and $n^{-}(k)$ are the action spectral densities of the surface and interfacial wave modes, 
respectively, and are related to $S^{+}(k)$ and $S^{-}(k)$ by $n^{+}(k)=S^{+}(k)/\omega^{+}(k)$ and $n^{-}(k)=S^{-}(k)/\omega^{-}(k)$.
The existence of two delta functions in the expressions for $dn^{+}_k/dt$ and $dn^{-}_k/dt$ implies the fact that
only resonant interactions contribute to the temporal evolution of the spectra.

For 2-D two-layer fluid system, two types of 3-wave resonances are known, called \lq\lq type-A" and \lq\lq type-B"
\cite{Choi2021}.
The type-A resonance is a resonance between two surface waves and one interfacial wave, and its resonance condition
is given by
\begin{equation}
\bm{k}_1^{+}=\bm{k}_2^{+}+\bm{k}_3^{-},\quad \omega_1^{+}=\omega_2^{+}+\omega_3^{-},
\label{eqn:type-A resonance condition}
\end{equation}
while the type-B is a resonance between one surface and two interfacial waves,
and its resonance condition is given by
\begin{equation}
\bm{k}_1^{+}=\bm{k}_2^{-}+\bm{k}_3^{-},\quad \omega_1^{+}=\omega_2^{-}+\omega_3^{-},
\end{equation}
where the superscripts $+$ and $-$ denote the surface and the interfacial wave modes, respectively.

When the wave propagation is restricted to 1-dimension as done here,
the type-A resonance contains the following two classes:
\begin{itemize}
\item
Class1: $\ds{k_1^{+}=-k_2^{+}+k_3^{-},\quad \omega_1^{+}=\omega_2^{+}+\omega_3^{-}}$,
\quad $(\theta_2^{+},\theta_3^{-})=(\pi,0)$,
\item
Class3: $\ds{k_1^{+}=k_2^{+}+k_3^{-},\ \omega_1^{+}=\omega_2^{+}+\omega_3^{-}}$,
\quad $(\theta_2^{+},\theta_3^{-})=(0,0)$.
\end{itemize}
while the type-B resonance contains the following two classes:
\begin{itemize}
\setlength{\itemsep}{0pt}
\item
Class2: $\ds{k_1^{+}=\pm k_2^{-} \mp k_3^{-},\ \omega_1^{+}=\omega_2^{-}+\omega_3^{-}}$,
 \quad $(\theta_2^{-},\theta_3^{-})=(\pi,0) \mbox{ or } (0,\pi) $,
\item
Class4: $\ds{k_1^{+}=k_2^{-}+k_3^{-},\ \omega_1^{+}=\omega_2^{-}+\omega_3^{-}}$.
\quad $(\theta_2^{-},\theta_3^{-})=(0,0)$.
%\ (possible only when $R\leq 1/3$)
\end{itemize}
Here $\theta_i\,(i=2,3)$ indicates the propagation direction of $k_i$ as seen from $k_1^{+}$,
and $k_j\,(j=1,2,3)$ denotes $|\bm{k}_j|$ and is positive-definite in these resonance conditions.
It is important to note that among these four resonance classes,
only Class 3 and Class 4 consist of waves propagating in the same direction.
However, according to \cite{Choi2021}, Class 4 is possible only when $R\leq 1/3$.
Therefore, under the parameter setting of our previous work where $R=0.8$,
Class3 resonance is the only possible type of resonance that would not generate waves propagating in the
opposite direction during the evolution in time.
This fact strongly suggests that the spectral evolution observed in our previous work
and shown again in Figs.\ref{fig:Spectra_R=0.8} and \ref{fig:Spectra_R=0.8_wide} of this paper
had been primarily brought about by the Class3 resonance.

\begin{figure}[h]
    \begin{center}
    \includegraphics[width=0.4\linewidth]
    {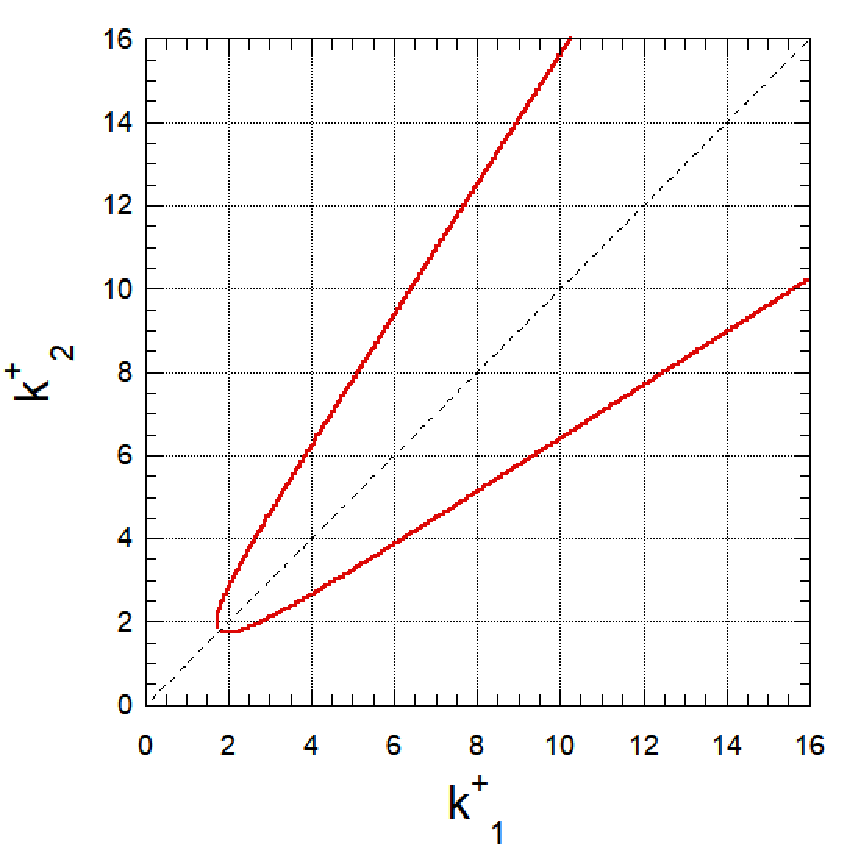}
    \end{center}
\caption{Wavenumbers of two surface waves satisfying the Class3 resonance condition ($h_l=2.0$, $R=0.8$)}
\label{fig:resonance curve}
\end{figure}

Figure \ref{fig:resonance curve} shows the wavenumbers of two surface waves satisfying the Class3 resonance
condition when $h_l=2.0$ and $R=0.8$.
%--------------------------------------------------------------------
%--------------------------------------------------------------------
It should be noted that there exists a minimum wavenumber $k_{\rm min}$ for surface waves participating in
the  Class3 resonance.  In the case shown, $k_{\rm min} \approx 1.73$.

\begin{figure}[h]
    \begin{center}
    \includegraphics[width=0.4\linewidth]
    {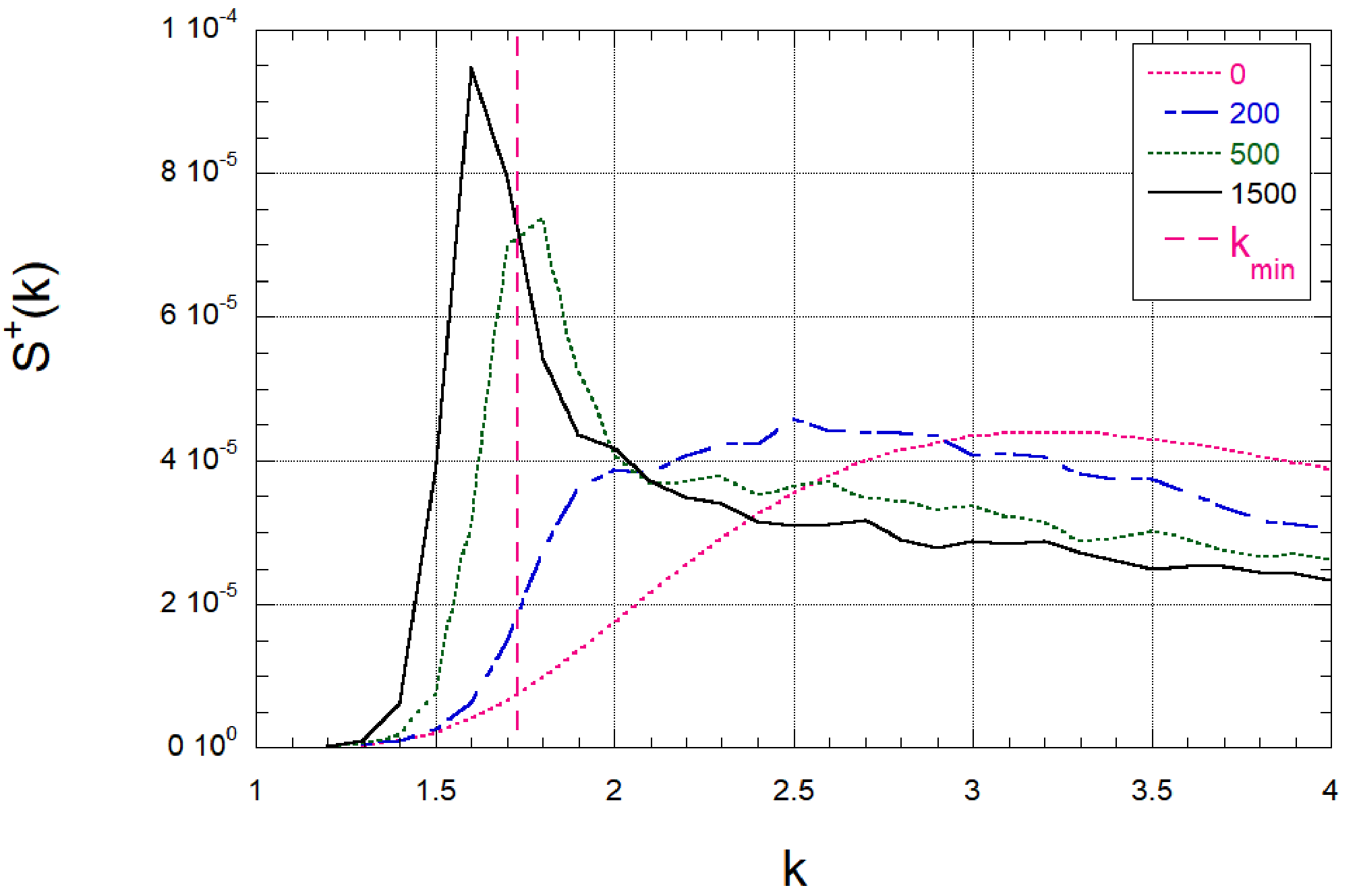}
\end{center}
\caption{\small Close-up view of Fig.~\ref{fig:Spectra_R=0.8}(a) around $k_{\rm min}$ }
\label{fig:S_spectrum_R=0.8_crit}
\end{figure}

Based on the discussion as above, it becomes apparent that there is a discrepancy between the spectral evolution
observed in our previous work and the standard WTT.
The conventional kinetic equations (\ref{eqn:kinetic equations}) based on the standard WTT indicates that
only resonant interactions contribute to the spectral evolution.
And the numerical fact that no waves propagating in the opposite direction were generated
during the spectral evolution as shown in Fig.(\ref{fig:Spectra_R=0.8_wide}) strongly suggests that
the resonant interactions playing the principal role in the spectral evolution are exclusively of Class3.
On the other hand, as illustrated in Fig.\ref{fig:resonance curve}, there exists a lower limit $k_{\rm min}$ in
the wavenumber of surface wave modes to participate in the Class3 resonance.  
Taking these facts together, it follows that the surface wave spectrum $S^{+}(k)$ should not change in time
below $k_{\rm min}$.
Nevertheless, as shown in Fig.~\ref{fig:S_spectrum_R=0.8_crit} which is an enlarged view of Fig.~\ref{fig:Spectra_R=0.8}(a)
 around $k_{\rm min}$, $S^{+}(k)$ actually changed significantly even in the range $k<k_{\rm min}\,(\approx 1.73)$.
%------------------------------------------------------------------
%------------------------------------------------------------------

\begin{figure}
%\hspace{10mm}
\begin{minipage}{0.5\linewidth}
\begin{center}
    \includegraphics[width=0.9\linewidth]
    {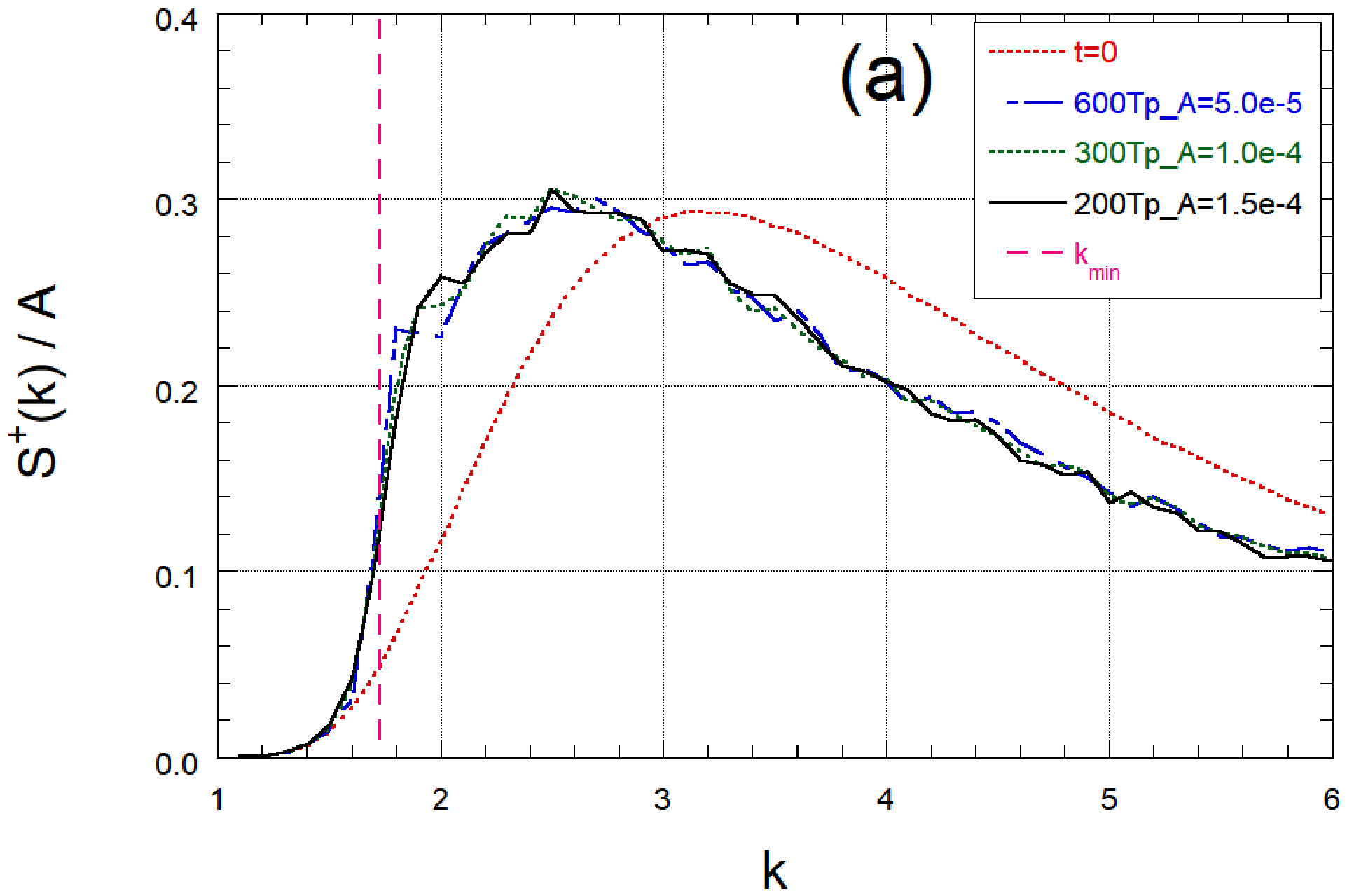}
\end{center}
\end{minipage}
%\hspace{0.05\linewidth}
\begin{minipage}{0.5\linewidth}
\begin{center}
    \includegraphics[width=0.9\linewidth]
    {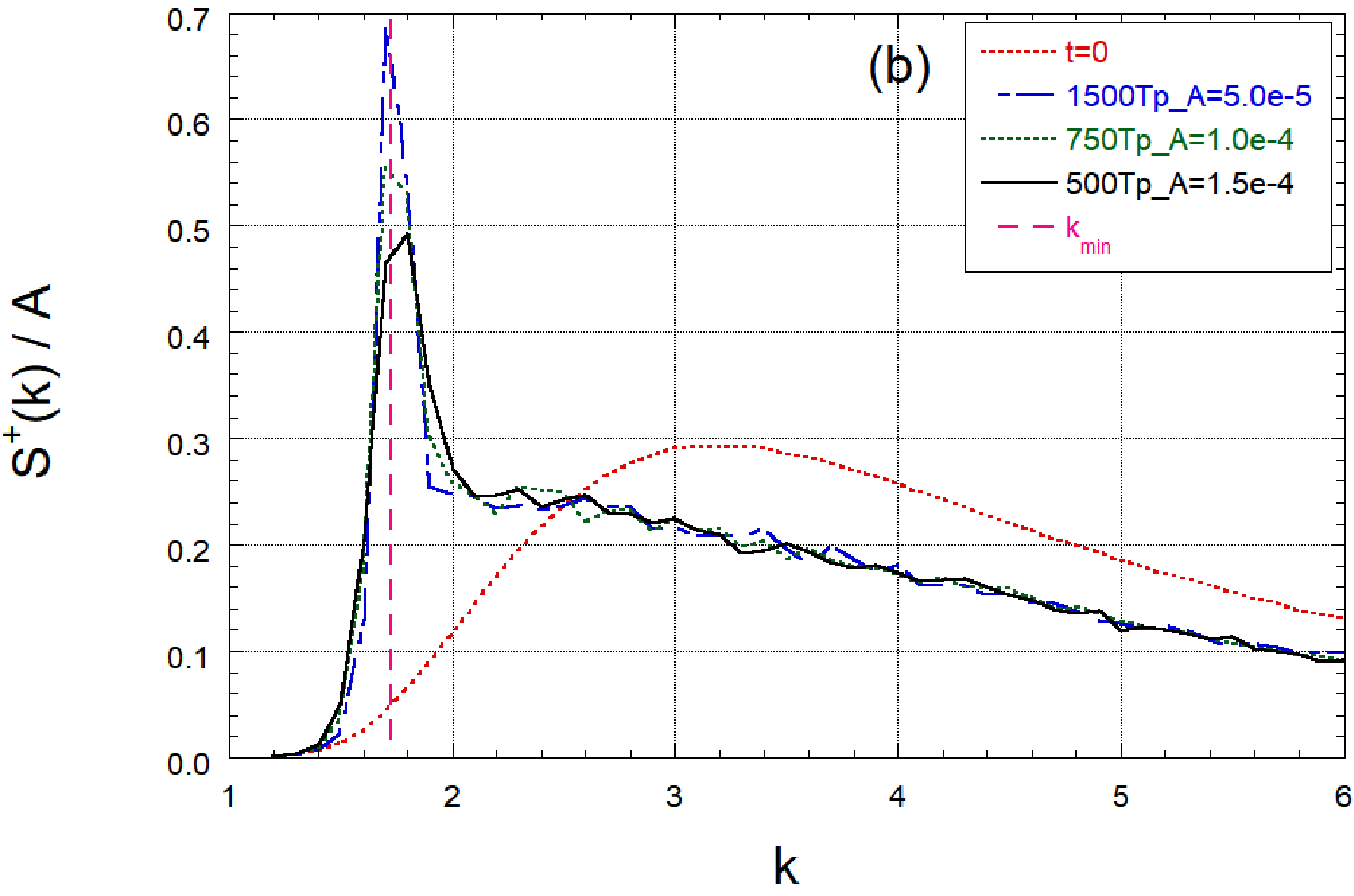}
\end{center}
\end{minipage}
\caption{\small Normalized surface wave spectra $\widetilde{S}^{+}(k)$ ($h_l=2$, $R=0.8$, $k_p=3.5$).
(a) $\tau=3\times10^{-2}$, (b) $\tau=7.5\times10^{-2}$.
}
\label{fig:3wave scaling_surface}
\end{figure}

The results of our previous work exhibited another characteristic that appeared to contradict the standard WTT, as shown below.
If the time evolution of $S^{+}(k)$ or $S^{-}(k)$ is truly governed by the standard kinetic equations (\ref{eqn:kinetic equations}), 
the normalized spectra $\widetilde{S}^{+}(k)\equiv S^{+}(k)/A$ and $\widetilde{S}^{-}(k)\equiv S^{-}(k)/A$,
where $A$ being the energy parameter $A$ in the initial spectrum (\ref{eq:initial spectrum}), 
should depend on $t$ and $A$ only through the \lq\lq slow time" $\tau\equiv A t$.
Figure~\ref{fig:3wave scaling_surface} shows $\widetilde{S}^{+}(k)$ obtained for three different values of $A$
in the case with $h_l=2$, $R=0.8$, $k_p=3.5$.
Figure~\ref{fig:3wave scaling_surface}(a) corresponds to three combinations of
($A$, $t$): ($0.5\times10^{-4}$, $600T_p$), ($1.0\times10^{-4}$, $300T_p$), and ($1.5\times10^{-4}$, $200T_p$),
all corresponding to the same scaled time $\tau=3.0\times10^{-2}$.
These correspond to a relatively early stage of spectral evolution, before the emergence of
the sharp peak in $S^{+}(k)$ around $k_{\rm min}$.
As expected, the three $\widetilde{S}^{+}(k)$ almost overlap, suggesting that $S^{+}(k)$ actually evolves according
to (\ref{eqn:kinetic equations}).
%------------------------------------------------------------------
%------------------------------------------------------------------
Figure~\ref{fig:3wave scaling_surface}(b) shows the same analysis for another three combinations:
($0.5\times10^{-4}$, $1500T_p$), ($1.0\times10^{-4}$, $750T_p$), and ($1.5\times10^{-4}$, $500T_p$),
all corresponding to $\tau=7.5\times10^{-2}$. 
These represent a later stage, after the emergence of a sharp peak in $S^{+}(k)$. 
While good agreement is still seen for most $k$, significant discrepancies appear around $k_{\rm min}$. 
This suggests that the formation of the sharp peak cannot be explained within the conventional framework
of WTT.
Furthermore, a closer look reveals that the time scale of peak formation is faster than the time scale of
spectral variation expected from the standard WTT, i.e., $O(1/\omega_0\epsilon^2)$, where
$\epsilon$ being the nondimensional amplitude and $\epsilon\propto \sqrt{A}$.
Incidentally, the term \lq\lq fast spectral evolution" in the title of this study refers to the spectral evolution
occurring at a timescale faster than that predicted by the standard WTT like that observed for $S^{+}(k)$
around $k_{\rm min}$.

%=====================================================================
\subsection{Breakdown of the Standard WTT Due to Double Resonance}
%\subsection{Tanaka (2018)}
%=====================================================================
%---------------------------------------------------------------------------------------------------

The resonance condition for Class 3 resonance is expressed as:
\begin{equation}
k_2^{+} = k_1^{+} + k_3^{-}, \quad \omega_2^{+} = \omega_1^{+} + \omega_3^{-}
\quad (k_2^{+} > k_1^{+}>0)
\label{eqn:Class3 resonance condition}
\end{equation}
On the resonance curve as shown in Fig.~\ref{fig:resonance curve}, $d k_1^{+} / d k_2^{+} = 0$ when $k_1^{+} = k_{\rm min}$.
Therefore, the derivative of (\ref{eqn:Class3 resonance condition}) with respect to $k_2^{+}$ 
at this point gives,
\begin{equation}
\frac{d \omega_2^{+}}{d k_2^{+}}
= \frac{d \omega_1^{+}}{d k_1^{+}}\frac{d k_1^{+}}{d k_2^{+}}
+ \frac{d \omega_3^{-}}{d k_3^{-}} \left( 1 - \frac{d k_1^{+}}{d k_2^{+}} \right)
= \frac{d \omega_3^{-}}{d k_3^{-}},
\end{equation}
implying the following fact:
Let $\left(k_1^{+}, k_2^{+}\,(> k_1^{+}), k_3^{-}\right)$ is a triad of Class3 resonance.
Then, if $k_1^{+}$ is equal to $k_{\rm min}$, 
the group velocities of the surface wave with $k_2^{+}$ and the interfacial wave
with $k_3^{-}$ are equal. %\cite{Tanaka_RIMS(2018)}
This fact has important implications for the kinetic equations as follows.

The kinetic equation (\ref{eqn:kinetic equations}a) for $n^{+}_k (= S^{+}(k)/\omega(k))$, 
after eliminating integrals using delta functions, can be rewritten as:
\begin{align}
\frac{d n^{+}_k}{d t} &=
-2 \pi
\frac{ |V^{(2)}_{k12}|^2 \{ (n^{+}_k - n^{+}_1) n^{-}_2 + n^{+}_k n^{+}_1 \} }{ \left| \frac{d \omega_1^{+}}{d k_1^{+}} - \frac{d \omega_2^{-}}{d k_2^{-}} \right| }
+ 2 \pi
\frac{ |V^{(2)}_{1k2}|^2 \{ (n^{+}_1 - n^{+}_k) n^{-}_2 + n^{+}_k n^{+}_1 \} }{ \left| \frac{d \omega_1^{+}}{d k_1^{+}} - \frac{d \omega_2^{-}}{d k_2^{-}} \right| },
\end{align}
where $k^{+}=k_1^{+}+k_2^{-}$ (hence $k^{+}>k_1^{+})$ in the first term on the RHS,
while $k^{+}=k_1^{+}-k_2^{-}$ (hence $k^{+}<k_1^{+})$ in the second term.
The denominator in the second term vanishes at $k = k_{\rm min}$, causing $d n^{+}_k / d t$ to diverge.
This indicates that the kinetic equation (\ref{eqn:kinetic equations}a) fails to correctly describe
the behavior of $S^{+}(k)$ at $k = k_{\rm min}$,
and the divergence of $d S^{+}(k)/d t$ toward infinity also suggests that $S^{+}(k)$ changes at a time scale
significantly faster than $O(1/\omega_0 \epsilon^2)$ expected from the standard WTT near $k = k_{\rm min}$.

On the other hand, the kinetic equation (\ref{eqn:kinetic equations}b) for the interfacial wave spectrum becomes
\begin{equation}
\frac{d n^{-}_k}{d t} = 2 \pi
\frac{ |V_{12k}|^2 \{ (n^{+}_1 - n^{+}_2) n^{-}_k + n^{+}_1 n^{+}_2 \} }{ \left| \frac{d \omega_1^{+}}{d k_1^{+}} - \frac{d \omega_2^{+}}{d k_2^{+}} \right| }.
\end{equation}
The denominator $\left| \frac{d \omega_1^{+}}{d k_1^{+}} - \frac{d \omega_2^{+}}{d k_2^{+}} \right|$ does not vanish
as long as $k_1^{+} \neq k_2^{+}$, and the kinetic equation remains well-defined.
This result is consistent with our previous observations (not shown here) that the time scale $O(1/\omega_0 \epsilon^2)$
was preserved for $S^{-}(k)$ even after the emergence of the sharp peak in $S^{+}(k)$ near $k_{\rm min}$.
The phenomenon that the kinetic equation governing the spectral evolution breaks down at points where
two group velocities of a resonant triad coincide has been known as \lq\lq double resonance".
The rapid growth of $S^{+}(k)$ near $k_{\rm min}$ we numerically observed in our previous study
is undoubtedly a manifestation of this double resonance phenomenon.

As far as we know, Benney \& Saffman \cite{Benney-Saffman} were the first to mention this situation and gave 
the name \lq\lq double resonance" to it. They pointed out that their method of derivation of the kinetic equation failed
in this situation, and proceeded with their discussion assuming that this situation did not occur.
Davidson\cite{Davidson1966, Davidson1967} also derived a kinetic equation for a system allowing three-wave resonances
with an entirely different technique than that of Benney \& Saffman.
There he also mentioned the double resonance and showed that, if the double resonance exists such that 
$\omega_k=\omega_{k'}+\omega_{k-k'}$ and $\frac{d}{dk'}(\omega_{k'}+\omega_{k-k'})=0$ hold simultaneously,
the \lq\lq collision integral" appearing in the kinetic equation would have secular parts growing as $t^{1/2}$ for large $t$.
(See Appendix ${\rm H}_2$ of \cite{Davidson1966} for more detail.)

Soomere\cite{Soomere1992, Soomere1993} investigated the double resonance problem in detail, focusing specifically
on Rossby waves.
In this case, the wavenumber is a two-dimensional vector.
Then the conditions for double resonance, i.e., $\bm{k}=\bm{k}_1+\bm{k}_2$, $\omega_{\bm{k}}=\omega_{\bm{k}_1}+\omega_{\bm{k}_2}$
and $\Delta\omega_{\bm{k}_1}=\Delta\omega_{\bm{k}_2}$ give five scalar conditions for six components of the three wavenumber
vectors, giving in general a one-dimensional set of solutions which he called the \lq\lq double resonance curve".
He also investigated the effect of the double resonance on the evolution of the whole spectrum and concluded that,
in spite of the fact that the kinetic equation fails in the vicinity of the set of double resonance points,
the error caused by its inability to correctly describe the energy transfer of (almost) double resonance Rossby waves is
small in comparison with the integral intensity of the energy transfer brought about by the infinite number of
ordinary (single) resonances.
He then suggested that the numerical study on the evolution of the Rossby wave spectrum may safely be carried out
based on the standard kinetic equation by just discarding the interactions within \lq\lq almost" double resonance triplets.

Amundsen\cite{Amundsen} and Amundsen \& Benney\cite{Amundsen-Benney} analytically calculated the asymptotic behavior
as $t\to \infty$ of the wave mode at the double resonance point deterministically based on a one-dimensional model equation
with quadratic nonlinearity,
and predicted that the amplitude $a(k)$ of the double resonance mode grows like $a(k) \propto \sqrt{t}$ 
and that the width of the peak around the mode narrows with $t$.

As seen above, the standard kinetic equation (\ref{eqn:kinetic equations}) breaks down when double resonances occur.
However, thanks to the previous researches such as Janssen \cite{Janssen}, Annenkov and Shrira \cite{Annenkov_Shrira2006},
and Gramstad and Stiassnie \cite{Gramstad-Stiassnie}, more generalized kinetic equations (GKE) are now available.
Unlike the standard kinetic equation, these GKE's account for the effects of non-resonant interactions as well as those
of resonant ones.
Therefore, they do not break down even when double resonances exist.
Furthermore, they do not assume that the spectrum changes only slowly.
For example, Annenkov and Shrira \cite{Annenkov_Shrira2015} applied the GKE to the study of rapid changes of
wind wave spectra caused by sudden wind speed changes such as squalls.
Consequently, it appears that these GKE's may be capable of adequately describing the emergence of a sharp peak
in $S^{+}(k)$ near the double resonance point $k_{\rm min}$ and its growth with a much faster rate than that
predicted by the standard WTT. 
%Examining this possibility is the objective of the present study.
The purpose of the present study is to verify whether this conjecture is indeed correct.

Incidentally, it may be worth mentioning that in plasma physics, the term \lq\lq double resonance" is also used but in a
completely different meaning in the context of wave-particle interactions (see \cite{Soto} for example.),
and some care is necessary not to confuse it with the phenomenon treated here.

%=====================================================================
\section{Simplified Two-Layer Fluid System and its DNS}
%=====================================================================
%=====================================================================
\subsection{Simplified Two-Layer Fluid System}
%=====================================================================
In order to achieve our aim as above, we need to derive the GKE for the two-layer fluid system.
According to the standard procedure of WTT, the derivation of the kinetic equation
(including GKE) starts from the Hamiltonian formalism of the system.

It has been shown by Choi et al.\cite{Choi2021} that the two-layer fluid system governed by
(\ref{eqn:governing equations}) can be expressed as the following Hamiltonian system under the approximation of
discarding all the interactions among four or more waves:
\begin{align}
\frac{da_k}{dt} &=-i\frac{\delta H}{\delta a_k^*}=-i\omega^{+}_k a_k-i\frac{\delta H_3}{\delta a_k^*}, \qquad
\frac{db_k}{dt}=-i\frac{\delta H}{\delta b_k^*}=-i\omega^{-} b_k -i\frac{\delta H_3}{\delta b_k^*} \nonumber\\
H&=H_2+H_3, \qquad H_2=\int \left(\omega^{+}(k) |a(k)|^2+\omega^{-}(k)|b(k)|^2\right)\,dk, \nonumber\\
H_3 &=\iiint\left[\left\{V^{(1)}_{123}(a_1^* a_2 a_3+{\rm c.c.})+V^{(2)}_{123}(a_1^* a_2 b_3+{\rm c.c.}) \right.\right.\nonumber\\
&+V^{(3)}_{123}(b_1^* a_2 a_3+{\rm c.c.})+V^{(4)}_{123}(b_1^* b_2 a_3+{\rm c.c.}) \nonumber\\
&+\left. V^{(5)}_{123}(a_1^* b_2 b_3+{\rm c.c.})+V^{(6)}_{123}(b_1^* b_2 b_3+{\rm c.c.})\right\}\,\delta^k_{1-2-3} \nonumber\\
&+\left\{V^{(7)}_{123}(a_1 a_2 a_3+{\rm c.c.})+V^{(8)}_{123}(a_1 a_2 b_3+{\rm c.c.})\right. \nonumber\\
&\left.\left.+V^{(9)}_{123}(a_1 b_2 b_3+{\rm c.c.})+V^{(10)}_{123}(b_1 b_2 b_3+{\rm c.c.})\right\}\delta^k_{1+2+3}\right]\,dk_{123}.
\label{eqn:Hamiltonian by Choi(2021)}
\end{align}
Here, $a(k)$ and $b(k)$ are the canonical complex amplitudes of the surface and the interfacial wave modes, respectively,
$V^{(i)}_{123}\,(i=1,\cdots,10)$ denote complicated coupling coefficients depending on three wave numbers
$k_1, k_2, k_3$.
Abbreviated notations such as $a_1=a(k_1)$, $V^{(i)}_{123}=V^{(i)}(k_1, k_2, k_3)$,
$\delta^k_{1-2-3}=\delta(k_1-k_2-k_3)$, $dk_{123}=dk_1 dk_2 dk_3$ have also been used.

If this Hamiltonian, which includes all types of  three-wave interactions, were to be used as it is, 
the numerical calculations and the derivation of the GKE would become extremely laborious.
Since we now know that the most important mechanism for the spectral evolution observed in our previous work
is the \lq\lq Class3" resonance and that all the Class3 resonant (and near-resonant) interactions are contained
in the second term of $H_3$, it would be reasonable to approximate the interaction Hamiltonian
$H_3$ by extracting only this part as
\begin{equation}
H_3\approx \iiint V^{(2)}_{123}(a_1^* a_2 b_3+{\rm c.c.})\,\delta^k_{1-2-3}\,dk_{123}.
\label{eqn:H3_simplified}
\end{equation}
Then, the corresponding evolution equations for the complex amplitudes become
\begin{subequations}
\label{eqn:simple model}
\begin{align}
&\frac{da_k}{dt}=-i\omega^{+}_k a_k -i\iint V^{(2)}_{012} a_1 b_2\,\delta^k_{0-1-2}\,dk_{12}
      -i\iint V^{(2)}_{102} a_1 b_2^* \,\delta^k_{1-0-2}\,dk_{12},
\label{eqn:simple model_a}
\\
&\frac{db_k}{dt}=-i\omega^{-}_k b_k -i\iint V^{(2)}_{120} a_1 a_2^\ast\,\delta^k_{1-2-0}\,dk_{12},
\label{eqn:simple model_b}
\end{align}
\end{subequations}
which form the governing equations for the \lq\lq simplified" two-layer fluid system we employ in this study.

Initially, we intended to use $V^{(2)}_{012}$ for the two-layer fluid system as is.
However, after looking at \cite{Choi2021}, we found that $V^{(2)}_{012}$ appears to be an overly complicated function,
so we gave up on using it in its original form.
Furthermore, as mentioned above, the purpose of this study is not to investigate the two-layer fluid system,
but to investigate the applicability of the GKE to systems with double resonances,
so there is no need to place excessive importance on the closeness to the two-layer fluid system.
Then, for further simplification, we will tentatively replace the original form of $V^{(2)}_{012}$ with
a much simpler one like $V^{(2)}_{012}=1$ or $V^{(2)}_{012}=\sqrt{|k_2|}$ here.
If this extremely simplified model can actually reproduce a similar spectral evolution as that we observed in 
the original two-layer fluid system as those shown in Figs.\ref{fig:Spectra_R=0.8} and \ref{fig:Spectra_R=0.8_wide},
we can use it as a tool to investigate the applicability of the GKE to the fast spectral changes induced by double resonance.

%=====================================================================
\subsection{DNS of the Simplified Model}
%=====================================================================
In order to check if the simplified model can reproduce a similar spectral evolution as the original two-layer fluid
system, we performed DNS based on the governing equation (\ref{eqn:simple model}) of the simplified model.
By discretizing the wavenumber space $k$ at equal intervals, we obtain the following system of ordinary differential equations
corresponding to (\ref{eqn:simple model}):
\begin{subequations}
\label{eqn:simple model_discrete}
\begin{align}
&\frac{da_k}{dt}=-i\omega^{+} a_k -i\sqrt{2\pi}\sum_{k_1}\sum_{k_2} V_{012}\,a_1 b_2\,\delta^k_{0-1-2}
      -i\sqrt{2\pi}\sum_{k_1}\sum_{k_2} V_{102}\,a_1 b_2^* \,\delta^k_{1-0-2}, \\
&\frac{db_k}{dt}=-i\omega^{-} b_k -i\sqrt{2\pi}\sum_{k_1}\sum_{k_2} V_{120}\,a_1 a_2^\ast\,\delta^k_{1-2-0},
\end{align}
\end{subequations}
where $\delta^k_{0-1-2}$ represents the Kronecker delta, and the subscripts 0, 1, and 2 correspond to $k$, $k_1$, and $k_2$, respectively.
The appearance of $\sqrt{2\pi}$ is due to the relationship between the continuous Fourier transform 
and the discrete Fourier transform. For more details on the discretization method, see \cite{Tanaka-Yokoyama2013}.
In (\ref{eqn:simple model_discrete}), we have omitted the superscript $(2)$ on $V_{012}$.
We shall use this notation henceforth.

In the DNS of the simplified model (\ref{eqn:simple model_discrete}), 
the evolution of $a_k(t)$ and $b_k(t)$ are traced by 4th-order Runge-Kutta method with a fixed time step $\Delta t$.
The convolution sums of the nonlinear terms are evaluated pseudo-spectrally by using FFT,
and the aliasing errors generated in this process are properly removed by using the so-called 3/2-rule.
The initial spectrum is given by (\ref{eq:initial spectrum}) with $k_p$ being fixed as $k_p=3.5$ throughout this study.
The value of the parameter $A$ was chosen so that the energy density of the initial wave field $E$,
which is defined as $E=\sum_k \omega^{+}|a_k(0)|^2$, would have some specified value.
In most computations, the wavenumber space is truncated at $k_{\rm max}=k_p\times5$,
and the number of nodes $n_x$ in the physical space is $n_x=2^{18}$.
%and some cases $2^{19}=524,288$ for higher accuracy.
When this setting is used, the wave field consists of some 350,000 weakly interacting wave trains.
The time step $\Delta t$ is typically chosen as $\Delta t=T_p/50$, where $T_p\approx3.36$ for the present parameter setting,
i.e., $h_l=2.0$, $R=0.8$ and $k_p=3.5$.
The spectra $S^{+}(k)$ and $S^{-}(k)$ are evaluated approximately as
\begin{equation}
S^{+}(k)=\sum_m \omega_m^{+} |a_m|^2 \big/\Delta k_S , \qquad S^{-}(k)=\sum_m \omega_m^{-} |b_m|^2 \big/\Delta k_S,
\end{equation}
where  $\Delta k_S$ is the resolution of the spectra and is set as $\Delta k_S=0.01$ in most cases,
and the sum with respect to the node number $m$ is taken over those nodes of the DNS which fall in the spectral bin
of width $\Delta k_S$ centered at $k$.
The spectra $S^{+}(k)$ and $S^{-}(k)$ are also averaged over ten realizations which differ each other only in the set 
of random numbers used to specify the initial phases of the component waves.

\begin{figure}[h]
\begin{minipage}{0.5\linewidth}
\begin{center}
    \includegraphics[width=0.9\linewidth]
    {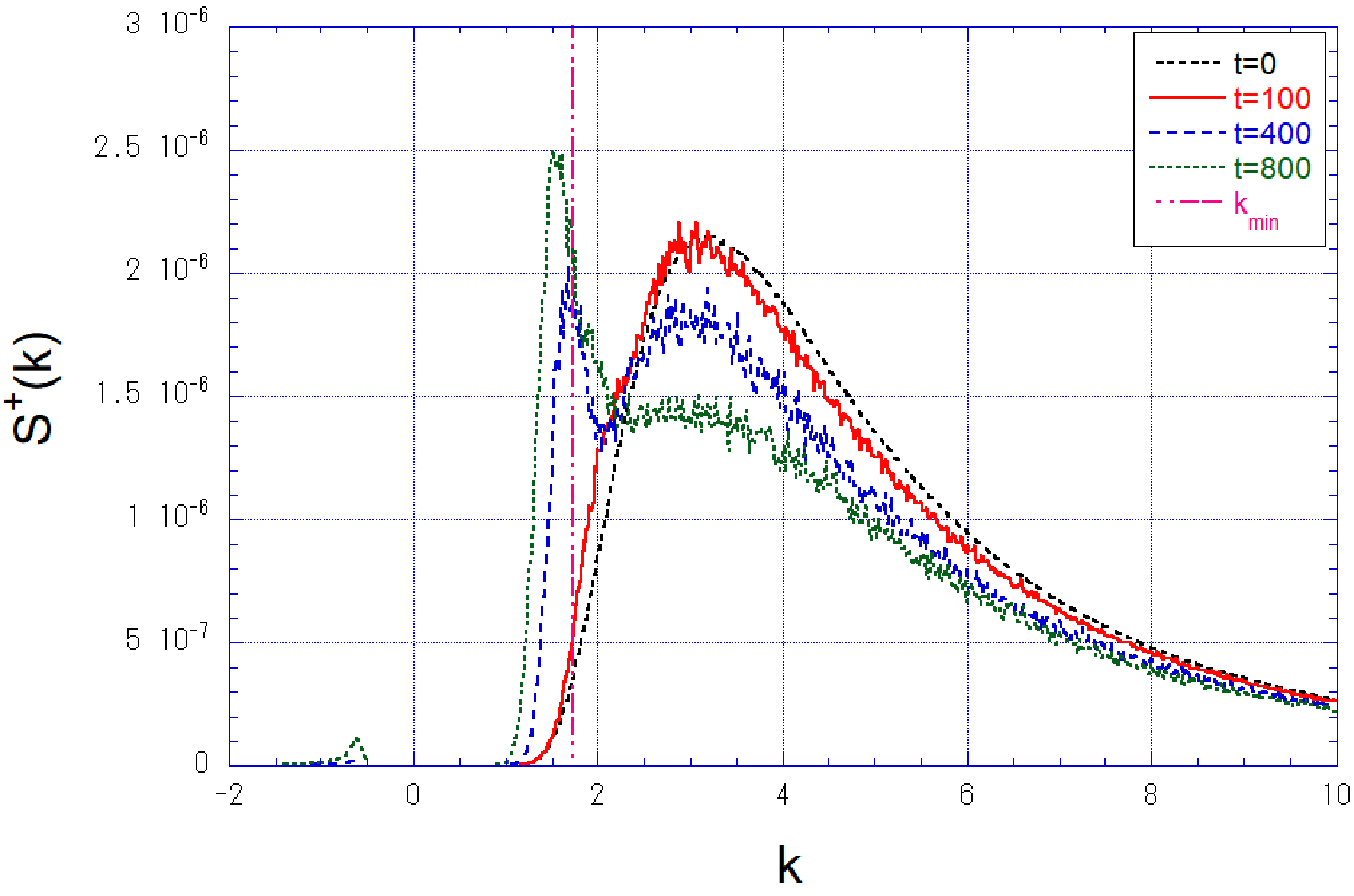}
\end{center}
\end{minipage}
%\hspace{0.1\linewidth}
\begin{minipage}{0.5\linewidth}
\begin{center}
    \includegraphics[width=0.9\linewidth]
    {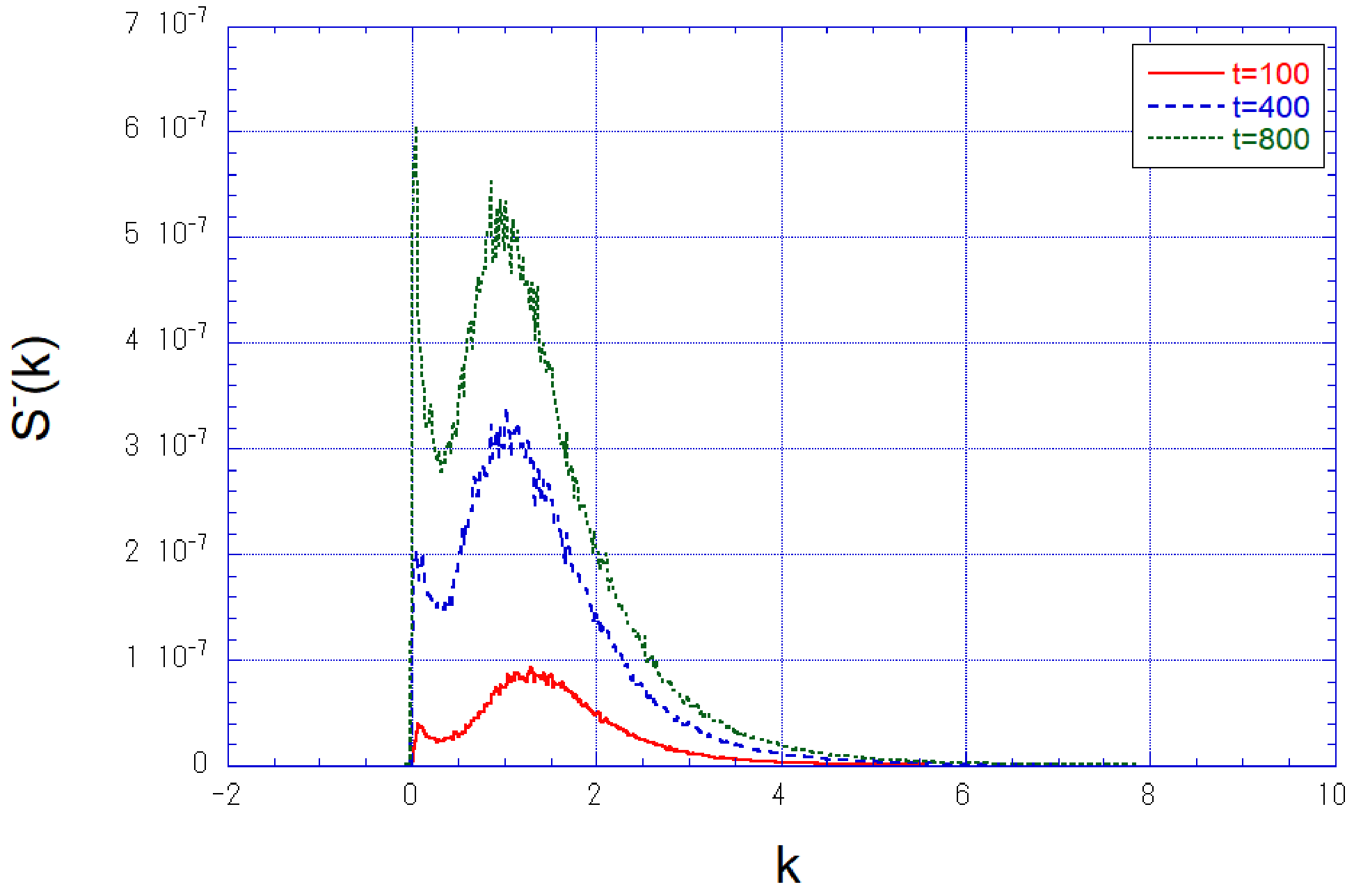}
    \end{center}
\end{minipage}
\caption{Example of evolution of $S^{+}(k)$ (left) and $S^{-}(k)$ (right) given by the DNS of the simplified model
with $V_{123}=1$ ($h_l = 2.0$, $R = 0.8$, $k_p = 3.5$, $E = 10^{-5}$)}
\label{fig:DNS_Spectra_V=1}
\end{figure}

Figure~\ref{fig:DNS_Spectra_V=1} shows an example of evolution of $S^{+}(k)$ (left) and $S^{-}(k)$ (right)
obtained by DNS of the simplified model with the choice $V_{012}=1$.
It can clearly be seen that, despite the significant simplifications,
the simplified model can reproduce similar behavior as the original two-layer fluid system
as far as $S^{+}(k)$ is concerned.

%------------------------------------------------------------------
%------------------------------------------------------------------

However, looking at the evolution of $S^{-}(k)$ shown on the right,
the simplified model turns out not to be so good.
As seen clearly in the figure, interfacial waves with very small wavenumbers grow rapidly.
This behavior is clearly undesirable from a numerical viewpoint,
and is also undesirable in the sense that such a behavior has not been observed in the DNS of the original two-layer fluid system.
This problem stems from our choice of $V_{123}$, i.e., $V_{123}=1$.
This $V_{123}$ does not vanish for the resonance of the type $k_1^{+} = k_2^{+}$, $k^{-} = 0$ which is called the
\lq\lq long-wave short-wave resonance". 
Due to the interactions close to this long-wave short-wave resonance, energy tends to constantly flow into the interfacial
waves with small wavenumbers, resulting in a sharp growth of $S^{-}(k)$ in the region $k^{-} \approx 0$ as observed
in Fig.\ref{fig:DNS_Spectra_V=1}.

\begin{figure}[h]
\begin{minipage}{0.5\linewidth}
\begin{center}
    \includegraphics[width=0.9\linewidth]
    {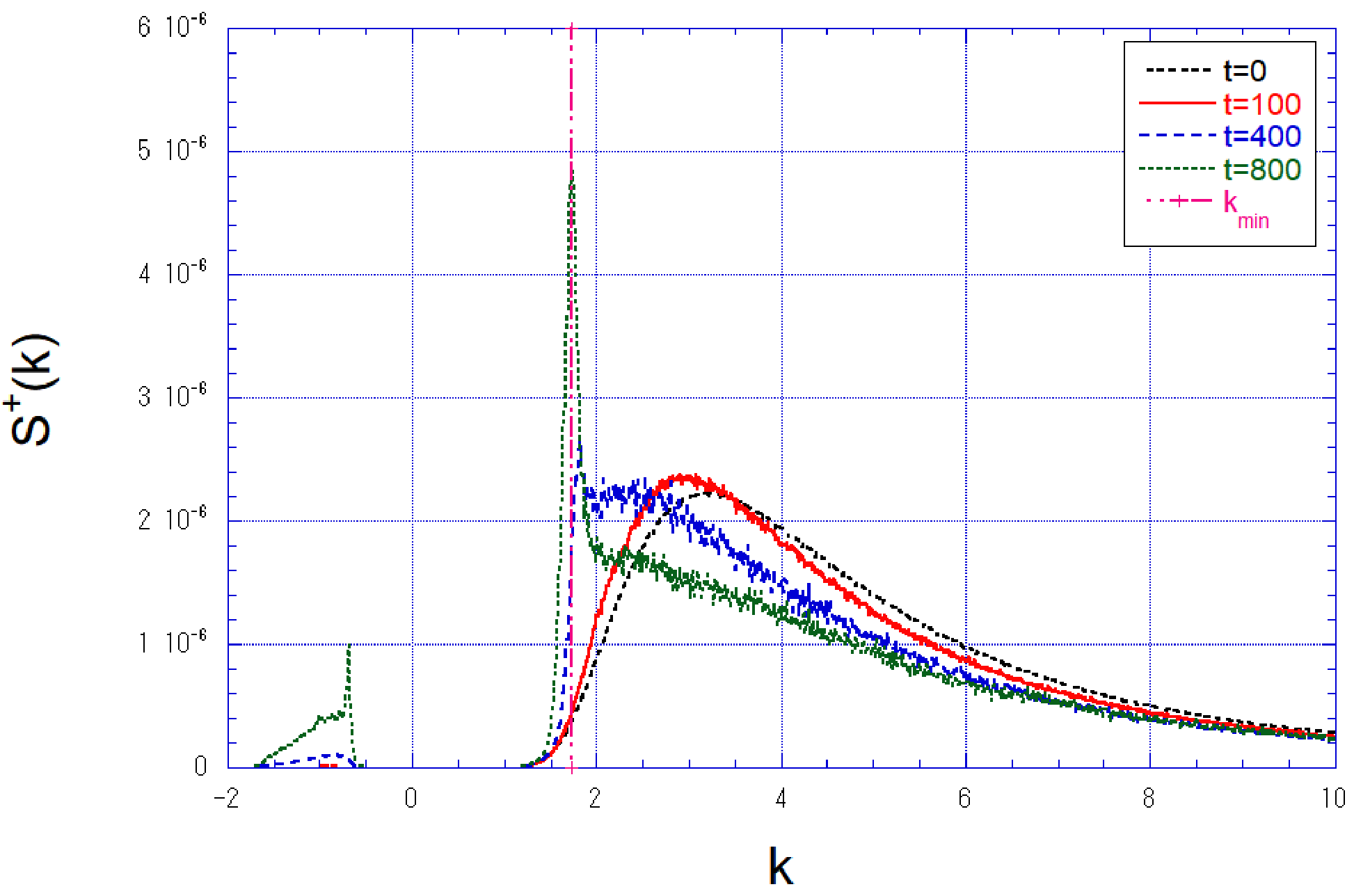}
\end{center}
\end{minipage}
%\hspace{0.1\linewidth}
\begin{minipage}{0.5\linewidth}
\begin{center}
    \includegraphics[width=0.9\linewidth]
    {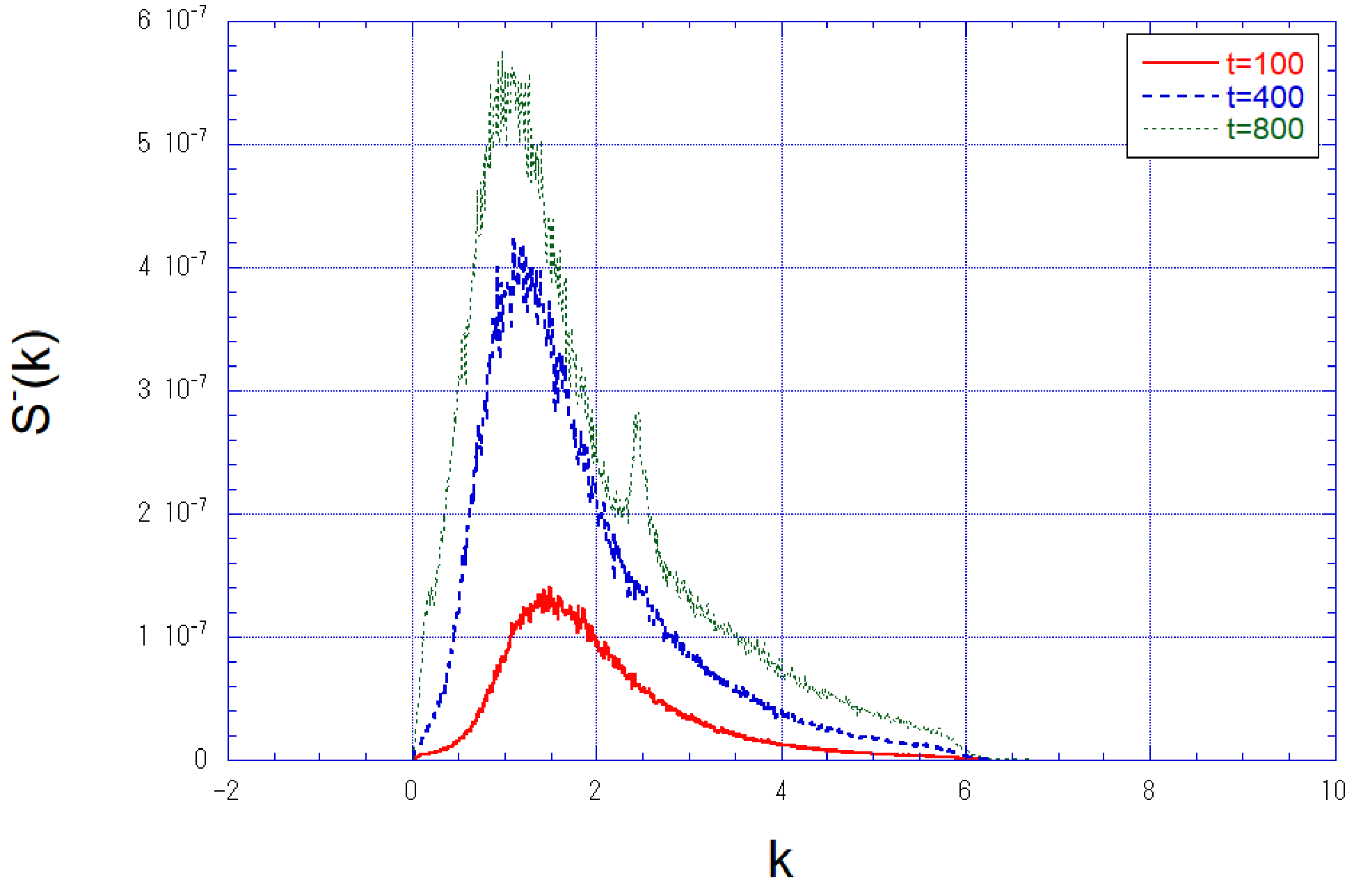}
    \end{center}
\end{minipage}
\caption{Same as Fig.\ref{fig:DNS_Spectra_V=1} but with $V_{123}=\sqrt{|k_3|}$ instead of $V_{123}=1$}
\label{fig:DNS_Spectra_V=k3^0.5}
\end{figure}

To circumvent this problem, we tried $V_{123}=\sqrt{|k_3|}$ instead of $V_{123}=1$, hoping that this $V_{123}$ would
eliminate the effect of the long-wave short-wave resonances.
The results are shown in Fig.\ref{fig:DNS_Spectra_V=k3^0.5}.

%------------------------------------------------------------------
%------------------------------------------------------------------

As the figure shows, this choice of $V_{123}$ successfully eliminates the unfavorable growth of $S^{-}(k)$ near $k=0$,
yet it still reproduces the growth of a sharp peak in $S^{+}(k)$ around $k_{\rm min}$.
So, we will employ this choice $V_{123}=\sqrt{|k_3|}$ in all the subsequent calculations in this work.
Using a larger power of $k_3$ for $V_{123}$ is also effective in eliminating the growth of $S^{-}(k)$ around $k\approx 0$,
but it would introduce at the same time a different problem of destabilization of the behavior of the spectra at higher wavenumbers.

\begin{figure}[!h]
%\hspace{10mm}
\begin{minipage}{0.5\linewidth}
\begin{center}
    \includegraphics[width=0.9\linewidth]
    {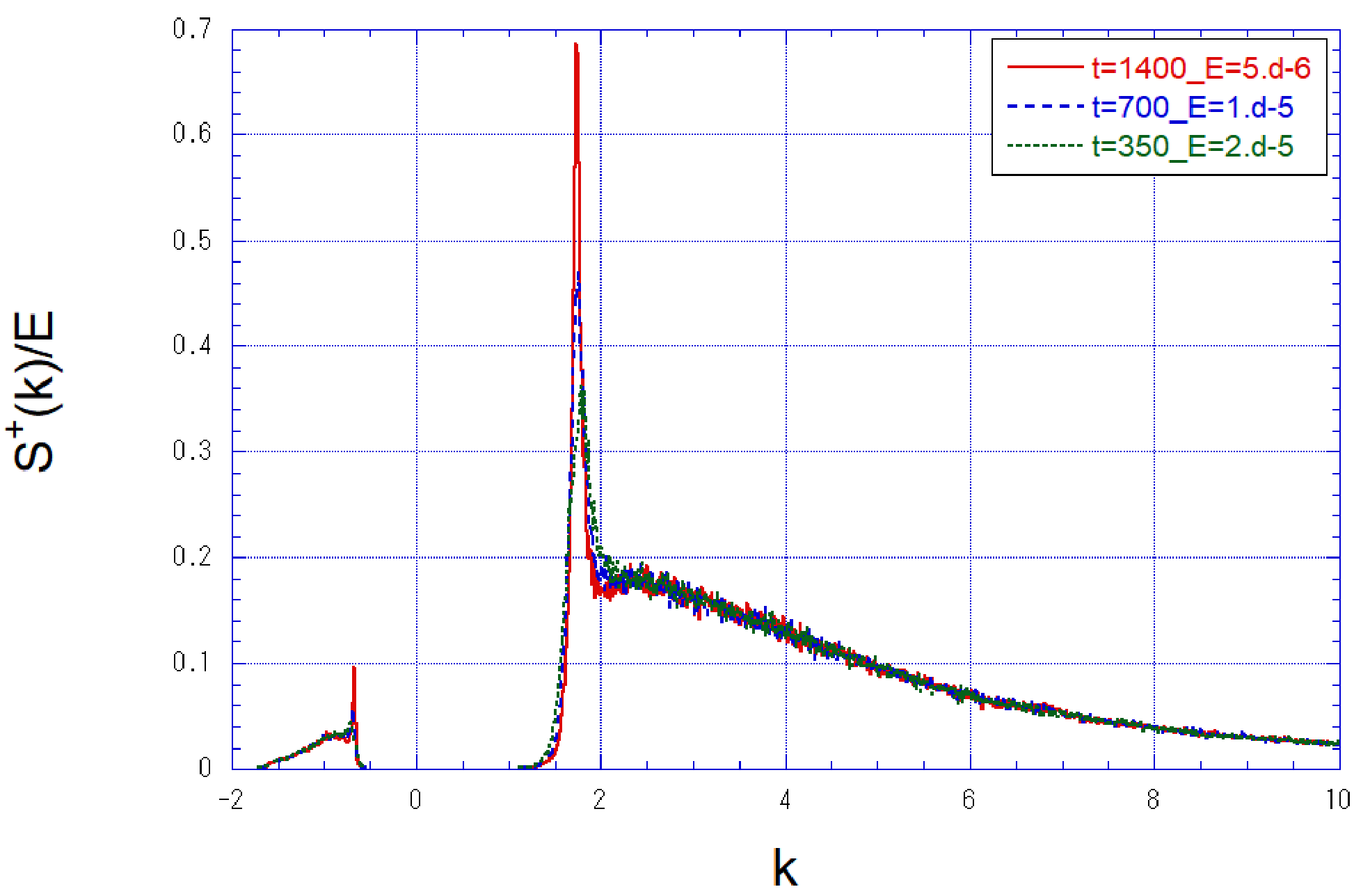}
\end{center}
\end{minipage}
%\hspace{0.05\linewidth}
\begin{minipage}{0.5\linewidth}
\begin{center}
    \includegraphics[width=0.9\linewidth]
    {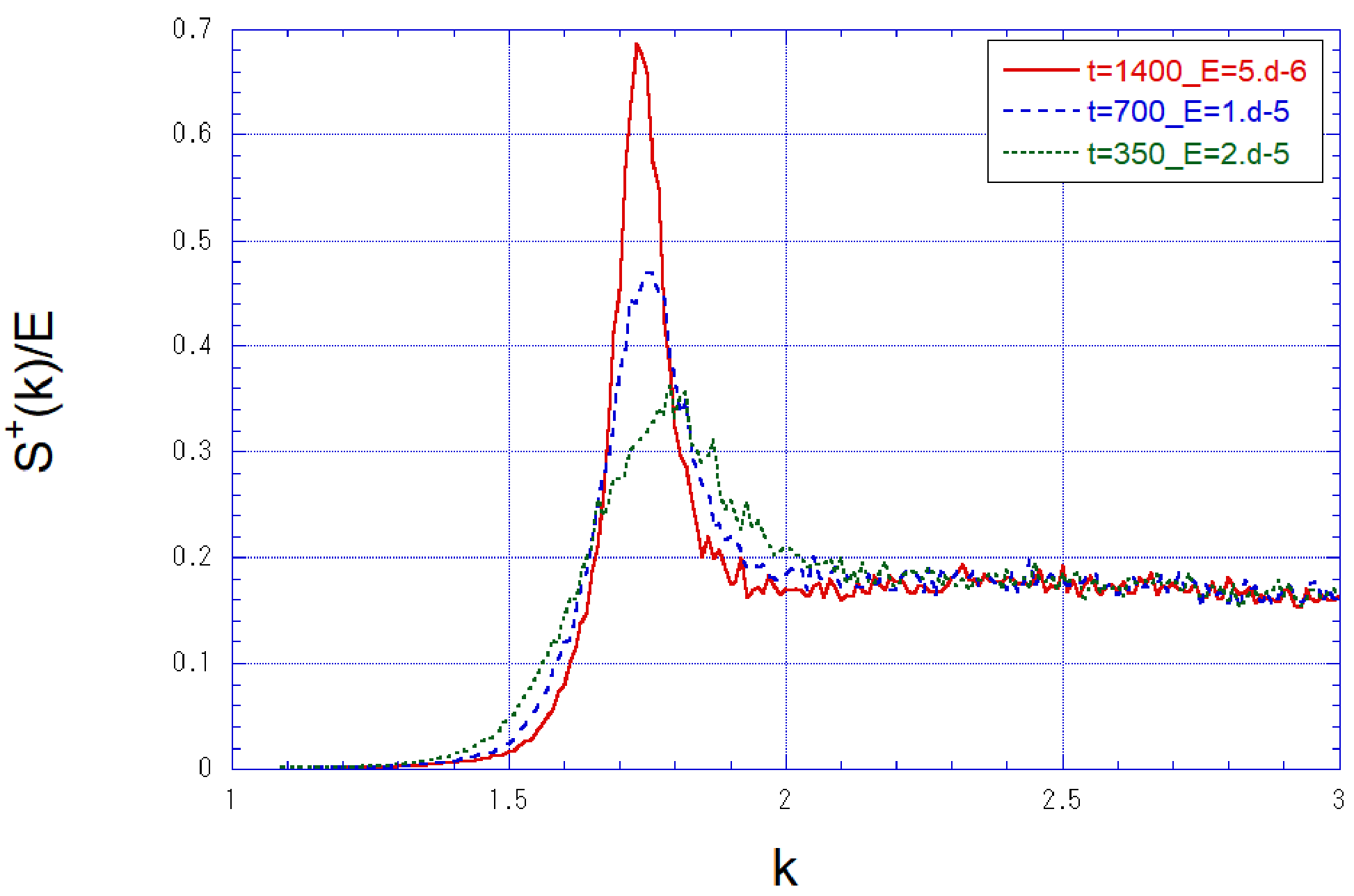}
\end{center}
\end{minipage}
\caption{Scaling character of the simplified model with respect to the slow time-scale $\tau (=Et)$.
The right figure is a closeup of the left one around $k_{\rm min}$.}
\label{fig:Scaling_SpectS_DNS_t=700_E=1.d-5}
\end{figure}

The simplified model with the choice $V_{123}=\sqrt{|k_3|}$ also has quite a similar time-scale characteristics to that
 of the original two-layer fluid system.
Figure \ref{fig:Scaling_SpectS_DNS_t=700_E=1.d-5} shows 
three $S^{+}(k,t)$'s with different combinations of $(E,t)$ all corresponding to the same value of $\tau\,(\equiv Et)$.
If the evolution of the spectrum $S^{+}(k)$ were to obey the standard kinetic equation (\ref{eqn:kinetic equations}),
the three $S^{+}(k)$'s should coincide for the whole range of $k$.
As Fig.\ref{fig:Scaling_SpectS_DNS_t=700_E=1.d-5} shows, the $S^{+}(k)$'s actually agree quite well for most of the range of $k$.
However, significant discrepancies among them appear in the region around the double resonance point $k_{\rm min}\,(\approx 1.73)$.
As clearly seen from the closeup shown on the right, the sharp peak appears to grow in a faster time-scale than $\tau$.
Comparison with Fig.\ref{fig:3wave scaling_surface}(b) reveals that these characteristics of the evolution of $S^{+}(k)$ given
by the simplified model are very similar to those observed previously for the original two-layer fluid system.
%-----------------------------------------------------------------
%------------------------------------------------------------------

All of the observations shown above strongly imply that the simplified model with the choice $V_{123}=\sqrt{|k_3|}$ can be
an appropriate tool for verifying the applicability of the GKE to a system with a double resonance.

%=============================
%=====================================================================
\subsection{Generalized Kinetic Equation for the Simplified Model}
%=====================================================================
As the statistical quantities to describe irregular wave fields, we introduce the second- and third-order moments defined as follows:
\begin{subequations}
\label{eqn:definition of n and J}
\begin{align}
&\langle a_k a_{k'}^\ast \rangle =n(k)\delta(k-k'),
\quad
\langle b_k b_{k'}^\ast \rangle =N(k)\delta(k-k'),\\
&
\langle a_0^\ast a_1 b_2 \rangle =J_{012}\delta(k-k_1-k_2).
\end{align}
\end{subequations}
Here, $\langle \quad \rangle$ denotes ensemble average, and the delta function arises from the assumption of spatial homogeneity of the field.

Differentiating (\ref{eqn:definition of n and J}a) with respect to $t$ and using (\ref{eqn:simple model}), we obtain
\begin{align}
&\frac{d n_k}{dt}=2 \,{\rm Im}\iint\left\{V_{012} J_{012} \delta_{0-1-2}
-V_{102} J_{102} \delta_{1-0-2} \right\}\,dk_{12},
\label{eqn:dn/dt}
\\
&\frac{d N_k}{dt}=-2\, {\rm Im}\iint V_{120}\, J_{120}\, \delta_{1-2-0}\,dk_{12}.
\label{eqn:dN/dt}
\end{align}

Differentiating the definition of $J_{012}$ in (\ref{eqn:definition of n and J}b) with respect to $t$ and
using (\ref{eqn:simple model}) and assuming weak non-Gaussianity so that the fourth-order cumulant
is negligible, we obtain the evolution equation for $J_{012}$ as follows:
\begin{equation}
\left\{\frac{d }{dt}-i\Delta\omega_{012}\right\}\,J_{012}=i V_{012} f_{012},
\quad
f_{012} \equiv n_1 N_2-n_0(n_1+N_2),
\label{eqn:dJ/dt}
\end{equation}
where $\Delta\omega_{012}=\omega_0-\omega_1-\omega_2$. 
Corresponding to the mode of $k_i$, $\omega_i\,(i=0,1,2)$ in $\Delta\omega_{012}$ should be replaced by
$\omega_i^{+}$ or $\omega_i^{-}$ properly.

The general solution of (\ref{eqn:dJ/dt}) is given by
\begin{equation}
J_{012}(t)=iV_{012}\int_0^t f_{012}(t'){\rm e}^{i\Delta\omega_{012}(t-t')}\,dt'
+J_{012}(0){\rm e}^{i\Delta\omega_{012}t}.
\label{eqn:general solution of J}
\end{equation}
Assuming here that
(i) $J_{012}(0)=0$, and that (ii)
the time variation of $n_k(t)$ and $N_k(t)$ (and hence that of $f_{012}(t)$) is much slower compared to the oscillation term
${\rm e}^{i\Delta\omega_{012}t}$, hence $f_{012}(t)$ can be treated as constant in the integral,
then (\ref{eqn:general solution of J}) gives
\begin{equation}
J_{012}(t)=iV_{012}\frac{{\rm e}^{i\Delta\omega_{012}t}-1}{\Delta\omega_{012}}.
\label{eqn:Janssen's J}
\end{equation}
Substituting this expression of $J_{012}(t)$ into (\ref{eqn:dn/dt}) and (\ref{eqn:dN/dt}) yields the 
\lq\lq Janssen's equation" \cite{Janssen}.

By assuming further in (\ref{eqn:Janssen's J}) that sufficient time has passed, the asymptotic evaluation as
$t\to\infty$ gives
\begin{equation}
J_{012}(t)=V_{012}\left[-\frac{\mathcal{P}}{\Delta\omega_{012}}+\pi i \delta(\Delta\omega_{012})\right].
\label{eqn:Hasselmann's J}
\end{equation}
Substituting this form of $J_{012}(t)$ into (\ref{eqn:dn/dt}) and (\ref{eqn:dN/dt}) yields 
the standard kinetic equations (\ref{eqn:kinetic equations}) of Zakharov-Hasselmann type.
\cite{Hasselmann1962,Nazarenko2011,Zakharov et al.2025}

Solving the system of equations (\ref{eqn:dn/dt}), (\ref{eqn:dN/dt}), and (\ref{eqn:dJ/dt}) to trace the time 
evolution of the spectra $n(k)$, $N(k)$, and $J_{012}$ constitutes 
the framework of the GKE for systems allowing three-wave resonant interactions as that treated here.
In GKE, it is also common to reduce the system of three equations to that of two equations involving only $n(k)$
and $N(k)$ by substituting (\ref{eqn:general solution of J}) into (\ref{eqn:dn/dt}) and
(\ref{eqn:dN/dt}) to delete $J$.  
However, in the present study, we do not adopt this reduction and instead directly handle the coupled system
of (\ref{eqn:dn/dt}), (\ref{eqn:dN/dt}), and (\ref{eqn:dJ/dt}).
This choice makes the numerical code much simpler, and it also makes it possible to handle directly the information
given by the third moment $J_{012}$.
Incidentally, we set $J_{012}(0)=0$ since the initial condition of our DNS consists of a superposition of a large number
of independent wavetrains with various wavenumbers for which $J_{012}$ should be zero.

When discretizing $k$ for numerical computation of the GKE of the simplified model,
both $n(k)$ and $N(k)$ are represented as one-dimensional arrays.
Although $J_{012}$ may appear at first sight to require a three-dimensional array due to its number of indices,
the delta function $\delta(k_0-k_1-k_2)$ in its definition (\ref{eqn:definition of n and J}b) allows us to
reduce it to a two-dimensional array.
If we denote $J_{012}\delta_{0-1-2}$ by a two-dimensional array $f(k,k_1)$, say,
then $J_{102}$ and $J_{120}$ can be expressed as $f(k_1,k)$ and $f(k_1,k_1-k)$, respectively.
Therefore, a single 2D array is sufficient to represent all information about $J$ appearing in the three evolution equations
of GKE.

The time integration of the discretized GKE system is performed by the fourth-order Runge-Kutta method with
a fixed time step $\Delta t$.
In integrating the evolution equation (\ref{eqn:dJ/dt}) for $J_{012}$,
we introduce an auxiliary variable $\widetilde{J}_{012}$ by $\widetilde{J}_{012}\equiv J_{012}\mathrm{e}^{-i\Delta\omega_{012}t}$
and applied the Runge-Kutta scheme to $\widetilde{J}_{012}$.
This allows us to use a larger time step $\Delta t$ without losing accuracy.
This technique has also been used in the DNS of (\ref{eqn:simple model}).
In most computations of GKE, the interval of $k$ is set as $\Delta k=0.01$ and the time step of RK4 is $\Delta t=T_p/10$.

%=====================================================================
\section{Results and Discussion: Comparison of DNS and GKE}
\begin{figure}[!h]
%\hspace{10mm}
\begin{minipage}{0.5\linewidth}
\begin{center}
    \includegraphics[width=0.9\linewidth]
    {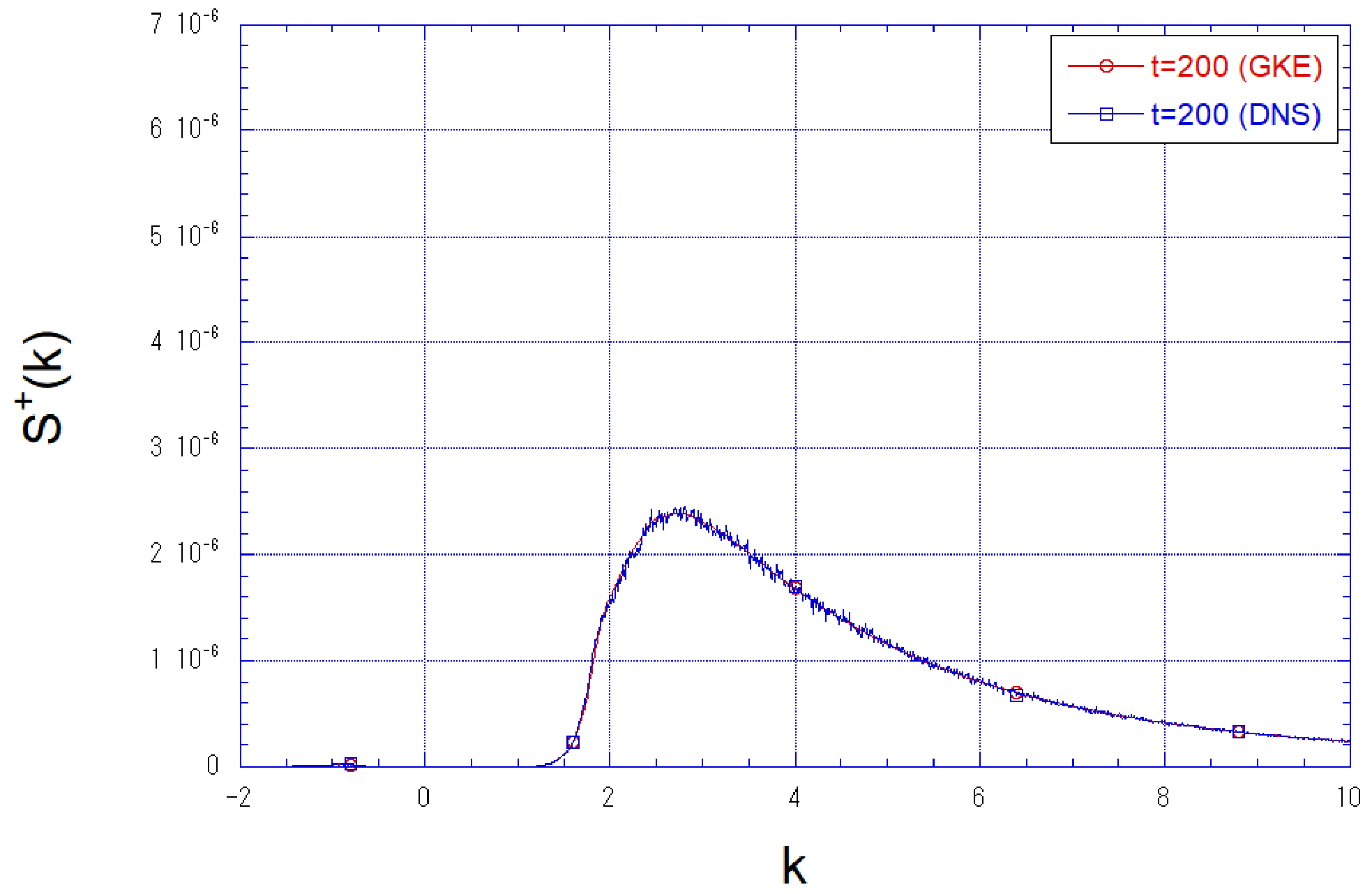}
\end{center}
\end{minipage}
%\hspace{0.05\linewidth}
\begin{minipage}{0.5\linewidth}
\begin{center}
    \includegraphics[width=0.9\linewidth]
    {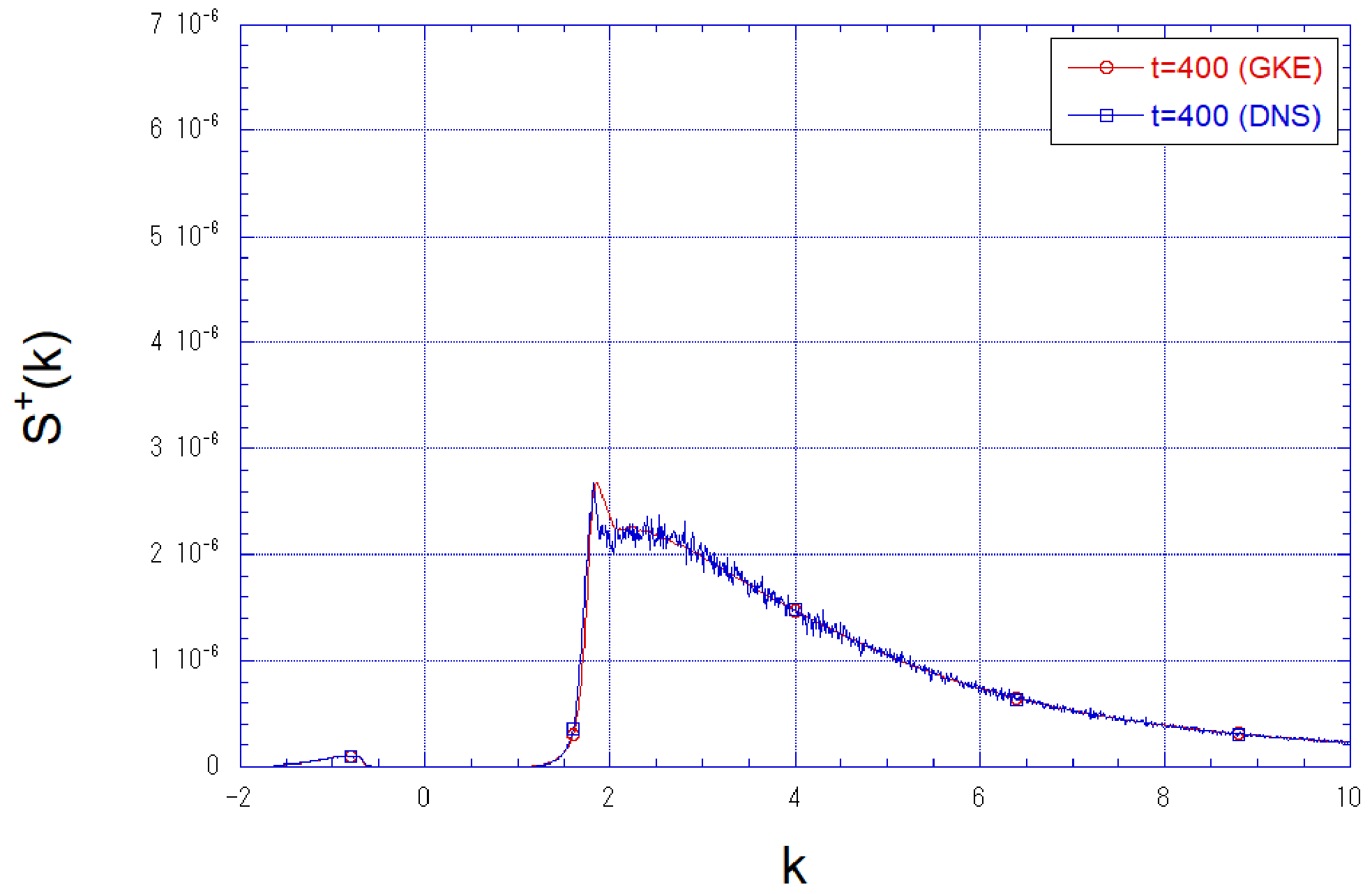}
\end{center}
\end{minipage}
%-----------------------------------------------------------------

\vspace{5mm}

%-----------------------------------------------------------------
\begin{minipage}{0.5\linewidth}
\begin{center}
    \includegraphics[width=0.9\linewidth]
    {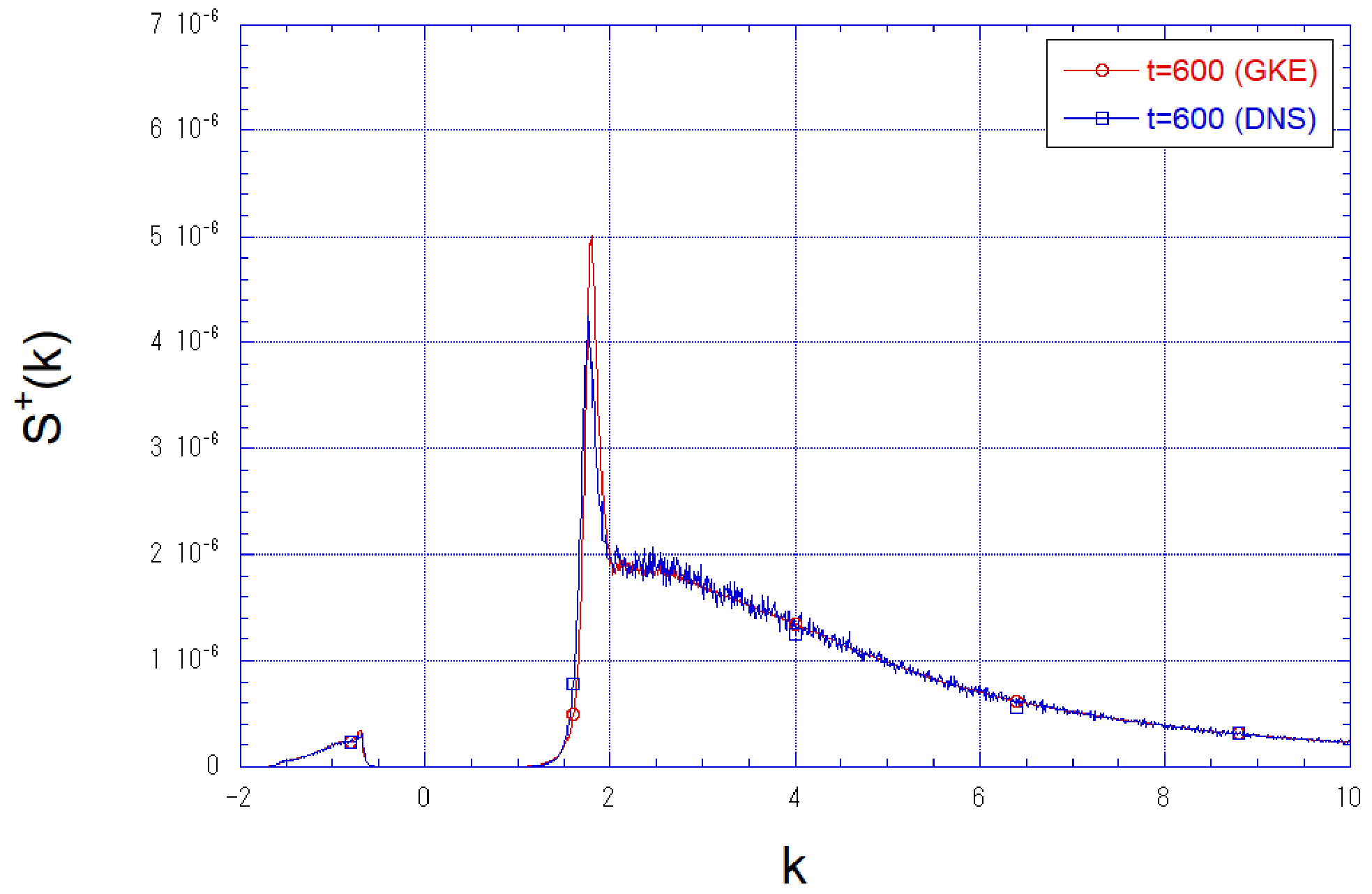}
\end{center}
\end{minipage}
%\hspace{0.05\linewidth}
\begin{minipage}{0.5\linewidth}
\begin{center}
    \includegraphics[width=0.9\linewidth]
    {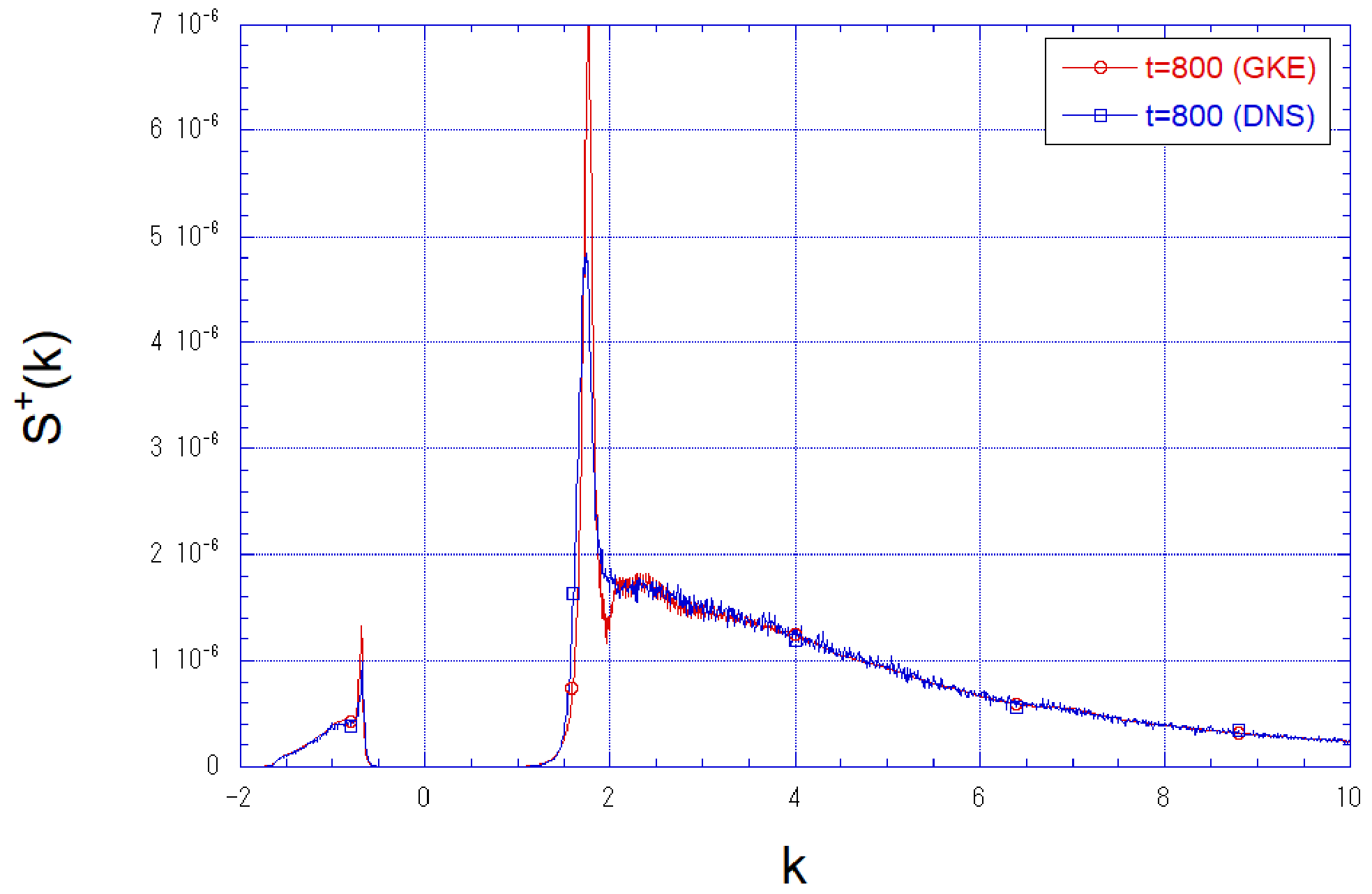}
\end{center}
\end{minipage}
\caption{Comparison of $S^{+}(k)$ obtained by DNS (blue line) and GKE (red line) of the simplified model
at different times: $200T_p$ (top left), $400T_p$ (top right), $600T_p$ (bottom left), $800T_p$ (bottom right).
$E=1\times10^{-5}$}
\label{fig:comparison_DNS_GKE_SpectS}
\end{figure}

%=====================================================================
%=====================================================================
%\subsection{Comparison between DNS and GKE of the Simplified Model}
%=====================================================================
Figure \ref{fig:comparison_DNS_GKE_SpectS} compares the energy spectrum $S^{+}(k)$ of the surface wave mode
obtained by the DNS (blue line) and the GKE (red line) for the simplified model at four different times, i.e.,
$200T_p$, $400T_p$, $600T_p$ and $800T_p$ when $E=1\times10^{-5}$.

%------------------------------------------------------------------
%------------------------------------------------------------------
At the early stage of the spectral evolution before the emergence of a sharp peak due to double resonance
such as $t=200T_p$ and $400T_p$, 
the results of GKE agree with those of DNS very closely. 
However, at later times, while the agreement still remains good except for the vicinity of the sharp peak
around the double resonance point $k_{\rm min}$,
there is a significant discrepancy between DNS and GKE near the peak. 
In particular, GKE tends to overemphasize the growth of the peak. 
Considering that the conventional kinetic equations (\ref{eqn:kinetic equations}) predict an infinite value of $dS^{+}(k)/dt$
at $k_{\rm min}$, it is a natural consequence that GKE, although the singularity at $k_{\rm min}$ is smeared out by including
the effects of non-resonant interactions, 
still tends to overestimate the growth of $S^{+}(k)$ in the region near $k_{\rm min}$.

The small wiggle seen in $S^{+}(k)$ given by DNS is probably due to an insufficient number of realizations (=10) used in
the estimate of the spectrum.
There is a noticeable dip in $S^{+}(k)$ given by GKE compared to that given by DNS around $k\approx 2$ when $t=800T_p$,
and this can be understood as follows.
Under the parameter setting used here, the wavenumbers of the other two waves that form a doubly Class3-resonant triad
with $k_1^{+}=k_{\rm min}\,(=1.73)$ are $k_2^{+}=1.97$ and $k_3^{-}=0.24$,
and this $k_2^{+}$ coincides with the position of the dip observed in $S^{+}(k)$ predicted by GKE.
As seen in Fig.\ref{fig:comparison_DNS_GKE_SpectS}, GKE tends to overestimate the growth of $S^{+}(k)$ at 
$k_1^{+}=k_{\rm min}$, which in turn likely to underestimate the energy of the surface wave at $k_2^{+}$ that resonates with
$k_{\rm min}$ hence supplies energy to support the excessive growth of the surface wave at $k_1^{+}=k_{\rm min}$.

Incidentally, the small peak appearing in the negative region at $k\approx -0.67$ can be understood as the surface wave
which resonates with the surface wave at $k_{\rm min}\,(=1.73)$ via the \lq\lq Class1 resonance".
Our interaction Hamiltonian (\ref{eqn:H3_simplified}), although greatly simplified from its original form, still contains
this Class1 resonance as well as the Class3 resonance which we are mainly concerned with.
The resonance conditions for the Class1 resonance are given by $k_1^{+}+ k_2^{+}=k_3^{-}$,
$\omega_1^{+}= \omega_2^{+}+\omega_3^{-}$,
and these conditions hold for $k_1^{+}=1.73$, $k_2^{+}=0.67$, and $k_3^{-}=2.40$.
Note that in this type of resonance the surface wave with $k_2^{+}$ propagates in the opposite direction from the other two waves.
By the selective growth of the surface mode at $k_1^{+}=1.79$ due to double resonance, the surface mode at $k_2^{+}=(-)0.67$
has also been generated via this Class1 resonance.

\begin{figure}[!h]
%\hspace{10mm}
\begin{minipage}{0.5\linewidth}
\begin{center}
    \includegraphics[width=0.9\linewidth]
    {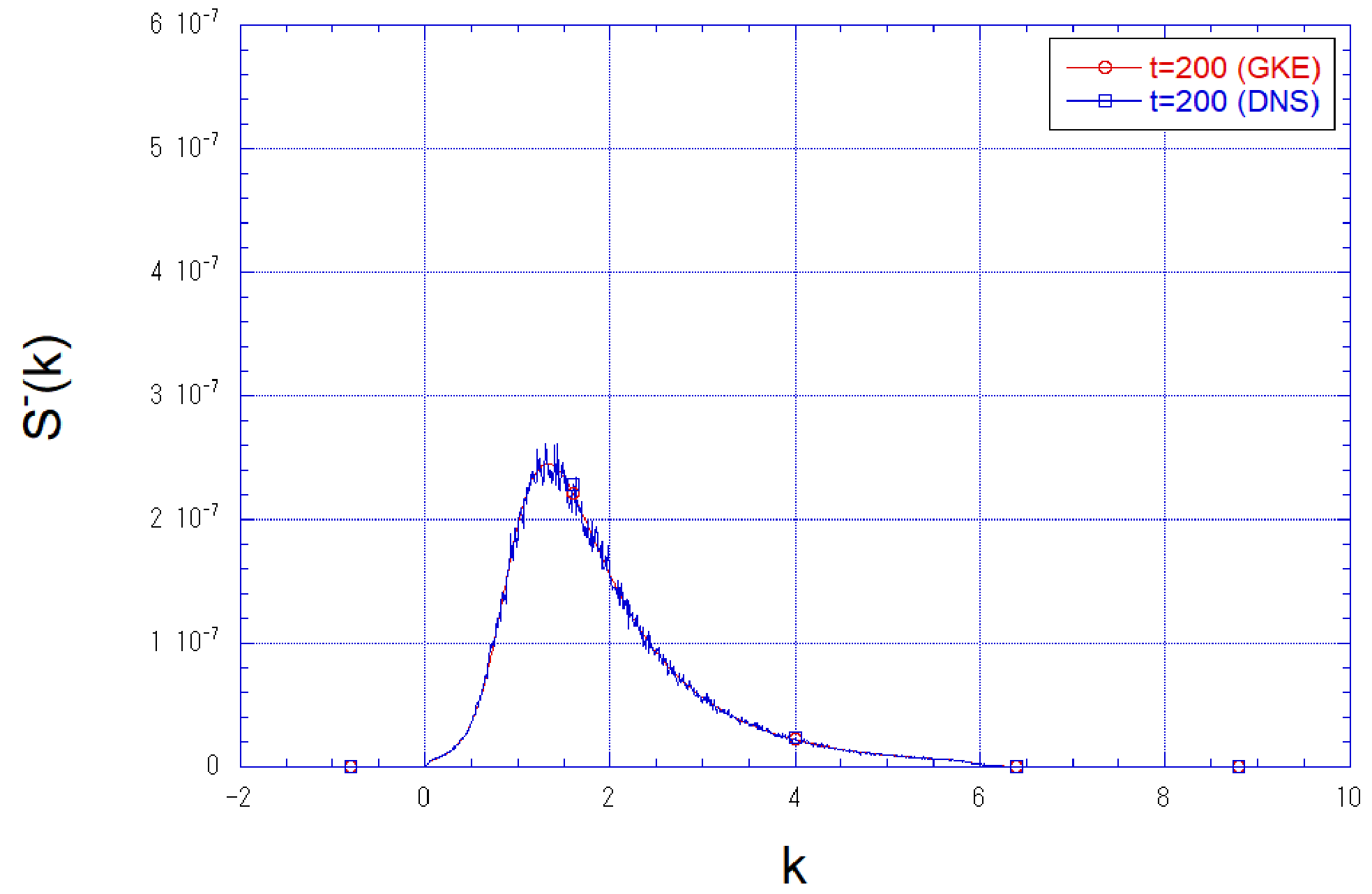}
\end{center}
\end{minipage}
%\hspace{0.05\linewidth}
\begin{minipage}{0.5\linewidth}
\begin{center}
    \includegraphics[width=0.9\linewidth]
    {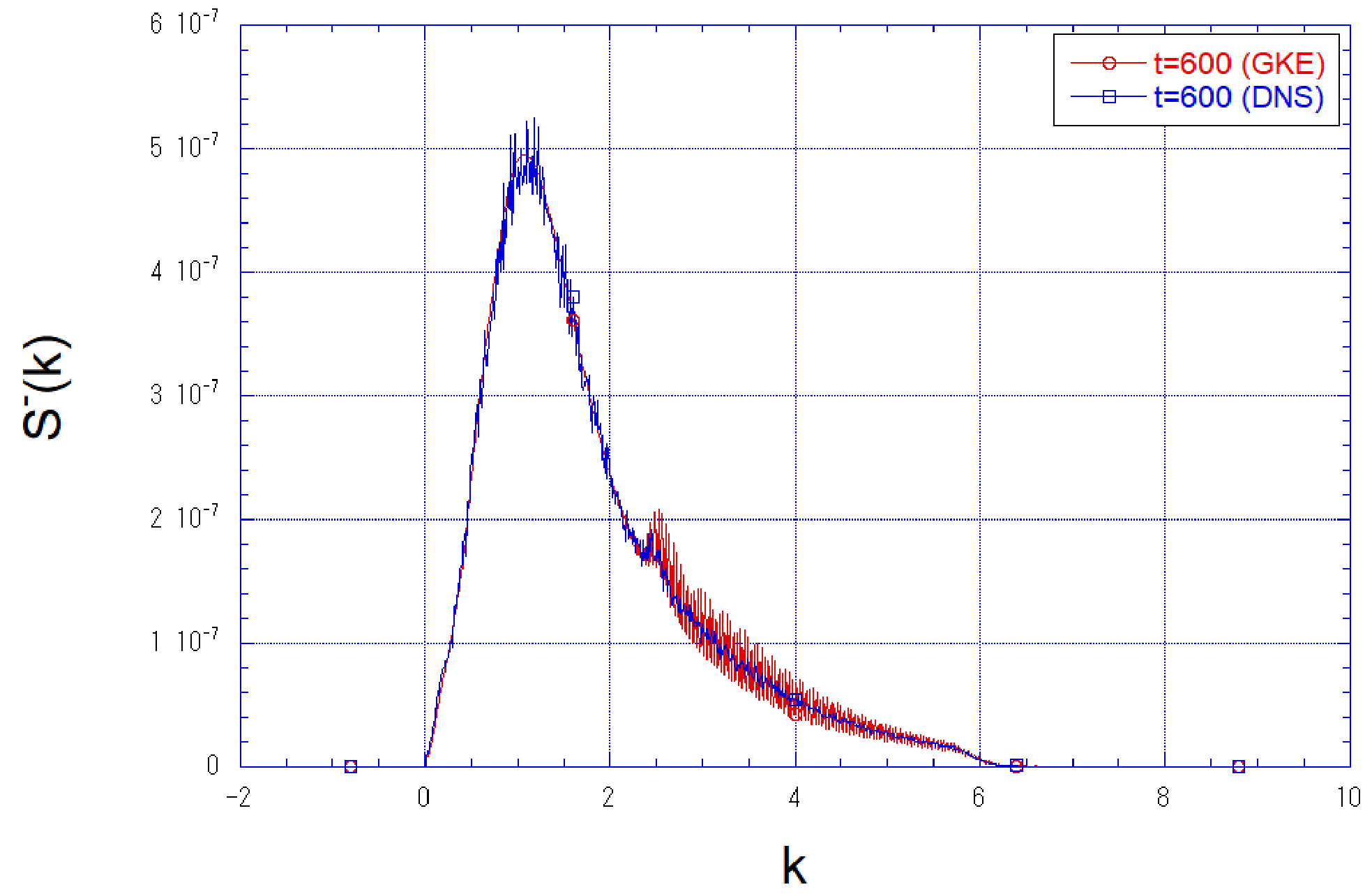}
\end{center}
\end{minipage}
\caption{Comparison of $S^{-}(k)$ obtained by DNS (blue line) and GKE (red line) of the simplified model at $t=200T_p$ (left) 
and $t=600T_p$ (right). $E=1\times10^{-5}$}
\label{fig:comparison_DNS_GKE_SpectI}
\end{figure}

Figure \ref{fig:comparison_DNS_GKE_SpectI} similarly compares the energy spectrum $S^{-}(k)$ of the interfacial
wave mode obtained by DNS (blue line) and GKE (red line) for the same case as that shown in Fig.\ref{fig:comparison_DNS_GKE_SpectS}.
For $S^{-}(k)$, unlike the comparison for $S^{+}(k)$ shown in Fig.\ref{fig:comparison_DNS_GKE_SpectS},
the agreement between DNS and GKE is excellent over the whole range of $k$ not only before the emergence of
the sharp peak in $S^{+}(k)$ due to double resonance ($t=200T_p$) but also after it has fully grown up ($t=600T_p$).
The slight hump observed around $k^{-}= 2.4$ both in DNS and GKE corresponds to a member of the resonant triad
of the Class1 resonance as that discussed above, and hence is another byproduct of the sharp growth of $S^{+}(k)$
around $k_{\rm min}$.

The weird oscillation in GKE observed beyond  $k^{-}= 2.4$ has probably been brought about by some kind of
numerical instability.  In the GKE computation shown in the figure, we discretized the $k$ space with the interval
$\Delta k=0.01$.  When we use a coarser discretization with $\Delta k=0.02$, the amplitude of the oscillation becomes
almost twice as large.  This suggests that the unfavorable oscillation would disappear if we use a sufficiently fine mesh.
Unfortunately, however, with our poor computing environment consisting of just one Windows PC,
this cannot be done, primarily due to insufficient memory.
%------------------------------------------------------------------
%------------------------------------------------------------------

\begin{figure}[!h]
%\hspace{10mm}
\begin{minipage}{0.5\linewidth}
\begin{center}
    \includegraphics[width=0.9\linewidth]
    {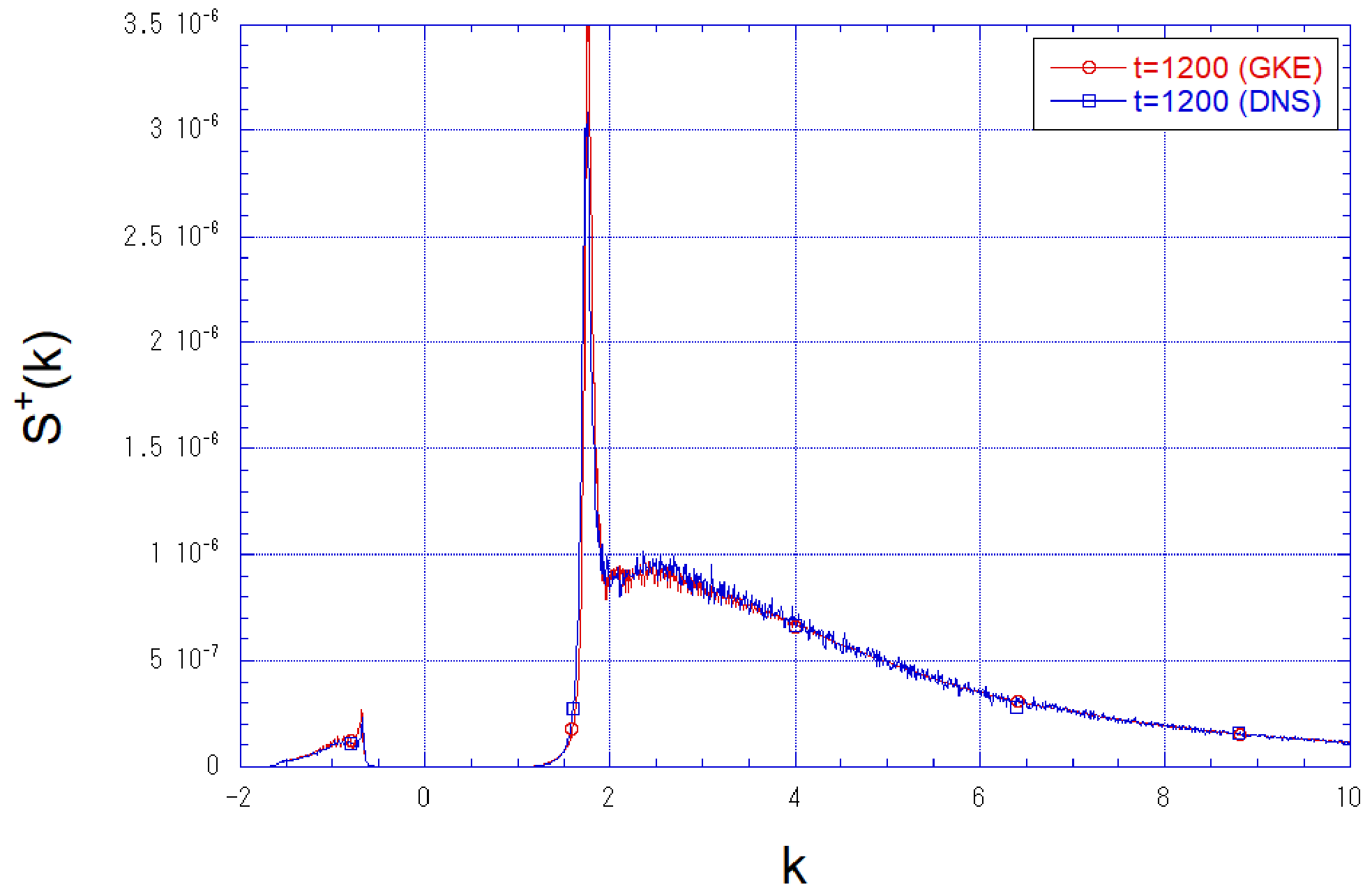}
\end{center}
\end{minipage}
%\hspace{0.05\linewidth}
\begin{minipage}{0.5\linewidth}
\begin{center}
    \includegraphics[width=0.9\linewidth]
    {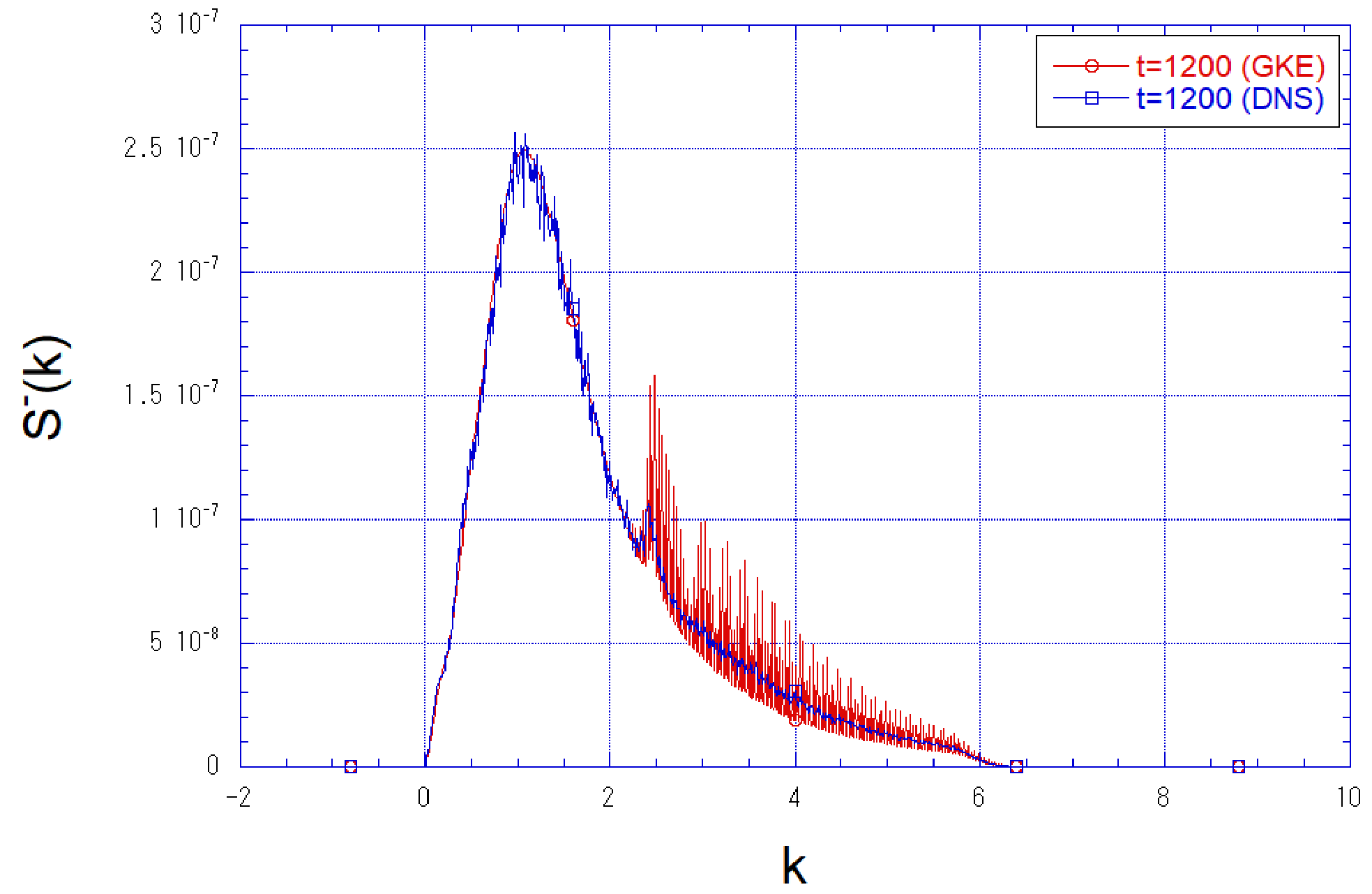}
\end{center}
\end{minipage}
\caption{$S^{+}(k)$ (left) and $S^{-}(k)$ (right) obtained by DNS (blue line) and GKE (red line) at $t=1200T_p$ when $E=5\times10^{-6}$.}
\label{fig:comparison_DNS_GKE_E=5e-6_t=1200}
\end{figure}

\begin{figure}[!h]
%\hspace{10mm}
\begin{minipage}{0.5\linewidth}
\begin{center}
    \includegraphics[width=0.9\linewidth]
    {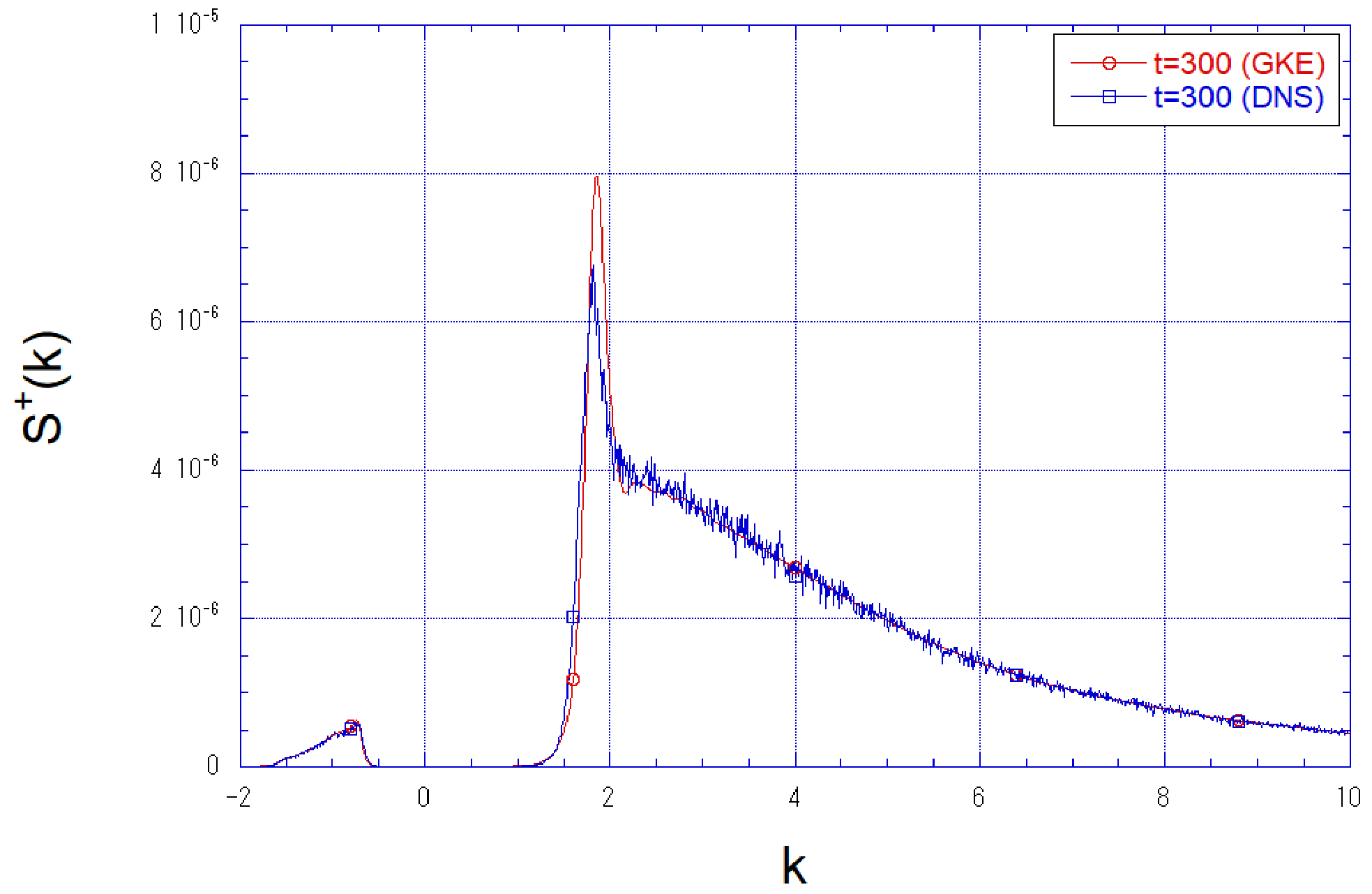}
\end{center}
\end{minipage}
%\hspace{0.05\linewidth}
\begin{minipage}{0.5\linewidth}
\begin{center}
    \includegraphics[width=0.9\linewidth]
    {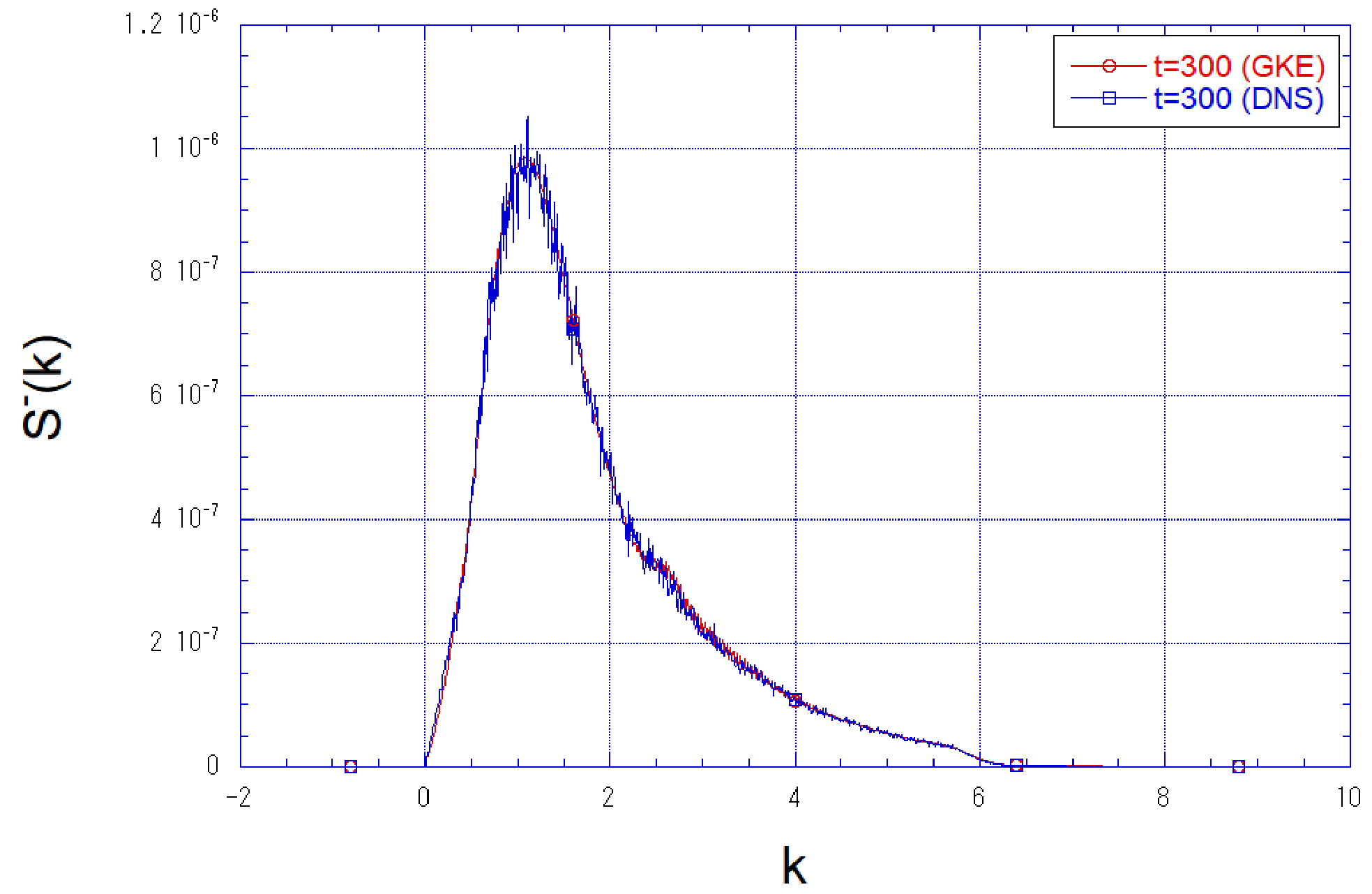}
\end{center}
\end{minipage}
\caption{Same as Fig.\ref{fig:comparison_DNS_GKE_E=5e-6_t=1200} but at $t=300T_p$ and $E=2\times10^{-5}$.}
\label{fig:comparison_DNS_GKE_E=2e-5_t=300}
\end{figure}

In order to see how the degree of nonlinearity of the wave field affects the applicability of GKE,
we carried out the computations with several different values of $E$. 
%Figs.\ref{fig:comparison_DNS_GKE_E=5e-6_t=1200} and \ref{fig:comparison_DNS_GKE_E=2e-5_t=300} show the results
%when $E=5\times10^{-6}$ and $E=2\times10^{-5}$, respectively.
Figure \ref{fig:comparison_DNS_GKE_E=5e-6_t=1200} shows $S^{+}(k)$ (left) and $S^{-}(k)$ (right) obtained by DNS (blue line)
and GKE (red line) at $t=1200T_p$ when $E=5\times10^{-6}$, while Fig.\ref{fig:comparison_DNS_GKE_E=2e-5_t=300} shows the
same but at $t=300T_p$ when $E=2\times10^{-5}$, both corresponding to the same value of the scaled time $\tau=Et$.
By comparing the $S^{+}(k)$'s shown in these figures, as well as that shown in the bottom left of
Fig.\ref{fig:comparison_DNS_GKE_SpectS} which shows $S^{+}(k)$ at $t=600T_p$ when $E=1\times10^{-5}$ hence also 
corresponding to the same value of $\tau$,
we can see that the peak in $S^{+}(k)$ around $k_{\rm min}$ caused by the double resonance gets more accentuated
as $E$ becomes smaller.  
This feature applies to both DNS and GKE.
Both DNS and GKE include non-resonant interactions of waves as well as resonant ones.
Although the spectral evolutions are primarily governed by the resonant interactions,
non-resonant interactions sufficiently close to resonance can also contribute to the spectral evolutions.
%due to frequency corrections arising from nonlinearity. 
In this case, the frequency mismatch $\Delta\omega$ allowed for those non-resonant interactions
that may affect the spectral evolution increases with $E$.
Consequently, the sharp peak in $S^{+}(k)$ appearing at the double resonance point loses its sharpness as $E$ increases,
spreading to the nearby wavenumber region.

Comparison of $S^{-}(k)$'s shown in Figs.\ref{fig:comparison_DNS_GKE_SpectI}, \ref{fig:comparison_DNS_GKE_E=5e-6_t=1200}, 
and \ref{fig:comparison_DNS_GKE_E=2e-5_t=300} clearly reveals that the numerical instability mentioned above in relation
to $S^{-}(k)$ in Fig.\ref{fig:comparison_DNS_GKE_SpectI} diminishes as $E$ increases,
and the unfavorable oscillation almost disappears when $E=2\times10^{-5}$.
As we mentioned above, the width of the sharp peak appearing in $S^{+}(k)$ due to double resonance becomes narrower
as $E$ decreases. Consequently, the need for a finer resolution in the $k$ space becomes more pronounced as $E$ decreases.
If we decrease $E$ while keeping the resolution $\Delta k$ fixed,
the numerical instability becomes increasingly severe as observed in the figures.

%------------------------------------------------------------------
%-----------------------------------------------------------------
%-----------------------------------------------------------------
%------------------------------------------------------------------

Incidentally, there still seems to be some room for debate regarding the magnitude of the frequency mismatch $\Delta\omega$
to be allowed.
In the case of surface gravity waves, for which three-wave resonance does not exist and four-wave resonance is the
lowest resonant interaction, it is commonly understood that $\Delta\omega \sim O(\epsilon^2)$.
However, there appear to be two different lines of reasoning that support this order estimate.
The first one is based on the amplitude of the \lq\lq instantaneous deviation" of the effective frequency $\Omega_k$ 
from the linear frequency $\omega_k$ as follows.
For four-wave interactions, the deterministic equations of motion can schematically be written as
\begin{equation}
\frac{da_k}{dt} = -i\omega_k a_k + i\sum_{k_1, k_2, k_3} V_{0123} a_1^\ast a_2 a_3\, \delta^k_{0+1-2-3}
\equiv -i\Omega_k a_k,
\label{eqn:schematic 4-wave}
\end{equation}
so the magnitude of instantaneous deviation of $\Omega_k$ from $\omega_k$,
which is given by the triple sum divided by $-i a_k$, is apparently of $O(\epsilon^2)$.
The second line of reasoning for $\Delta\omega \sim O(\epsilon^2)$ is based on the \lq\lq average deviation" of 
$\Omega_k$ from $\omega_k$ as follows.
Dividing the nonlinear term in (\ref{eqn:schematic 4-wave}) into two parts, it can be expressed as
\begin{equation}
\frac{da_k}{dt} = -i \omega_k a_k + i a_k \sum_{k_1} V_{0101} |a_1|^2 \, \delta^{k}_{0+1-0-1} 
+ i\sum_{\substack{k_1,k_2,k_3\\ k_1 \neq k_2,k_3}} V_{0123} a_1^\ast a_2 a_3 \, \delta^{k}_{0+1-2-3},
\end{equation}
where the average deviation of $\Omega_k$ from $\omega_k$ arises from the first sum and is of the order of $O(\epsilon^2)$ again.
Thus, for four-wave systems, both lines of reasoning lead to the same conclusion that $\Delta \omega$ is of order $O(\epsilon^2)$.

On the other hand, in three-wave systems like the one treated here,
the deterministic equations of motion can schematically be written as
\begin{equation}
\frac{da_k}{dt} = -i\omega_k a_k + i\sum_{k_1, k_2} V_{012} a_1 a_2\, \delta^k_{0-1-2},
%= -i\Omega_k a_k,
\label{eqn:schematic 3-wave}
\end{equation}
and the two arguments give different estimates for the magnitude of $\Delta \omega$.
The former reasoning based on the \lq\lq instantaneous" deviation of $\Omega_k$ from $\omega_k$ predicts
that $\Delta \omega$ is $O(\epsilon)$, while the latter reasoning based on the \lq\lq average" deviation
of $\Omega_k$ from $\omega_k$ would predict that $\Delta \omega$ is $O(\epsilon^2)$ as theoretically shown by
Jakobsen and Newell \cite{Jakobsen-Newell2004}.
In this sense, three-wave systems can provide more suitable platforms than four-wave systems for elucidating
which of the two reasonings concerning the order of magnitude of $\Delta \omega$  is more appropriate.
In any case, for this discussion, a systematic study focusing on simpler three-wave systems that do not involve
unnecessary complications such as double resonance would be more appropriate.

Before concluding this section, we briefly report a comparison between our numerical results and
the theoretical prediction by Amundsen\cite{Amundsen} and Amundsen\&Benney\cite{Amundsen-Benney}.
Amundsen\cite{Amundsen} and Amundsen\&Benney\cite{Amundsen-Benney} studied analytically the growth
of the mode at the double resonance point by using a simple 1-dim model equation with a quadratic nonlinearity.
Their approach is fully deterministic and the object of their analysis is the deterministic evolution of
the complex amplitudes of Fourier modes instead of statistical quantities such as the spectrum.
Their model equation is
\begin{equation}
u_t+\mathcal{L}(u)=\epsilon\,\mathcal{N}(u), \qquad u(x,0)=f(x),
\end{equation}
which is expressed in terms of the Fourier transform as
\begin{equation}
a_t=\epsilon\int H(l,k-l)a(l)a(k-l)\mathrm{e}^{-i\Delta\omega(k,l)t}\,dl, \quad a(k,0)=\hat{a}(k),
\end{equation}
where $\Delta\omega(k,l)=\omega(l)+\omega(k-l)-\omega(k)$,
and $H(l,k-l)$ represents the kernel of nonlinear interactions corresponding to the operator $\mathcal{N}$,
and $\hat{a}(k)$ is the Fourier transform of the initial function $f(x)$.

Expanding $a(k)$ in a perturbation series as $a(k,t)=a_0+\epsilon a_1+\cdots$,
they showed that $(a_0)_t=0$, i.e., $a_0(k)=\hat{a}(k)$, and $a_1(k)$ is given by
\begin{equation}
a_1(k)\approx \frac{2\sqrt{2\pi}\mathrm{e}^{-i\pi/4}H(\tilde{l},k-\tilde{l})a_0(\tilde{l})a_0(k-\tilde{l})}
{\sqrt{\Delta''\omega(k,\tilde{l})}}
\sqrt{t} \int_0^1 \mathrm{e}^{-\Delta\omega(k,\tilde{l})tv^2}\,dv.
\label{eqn:a1(k)}
\end{equation}
Here $\tilde{l}$ is the wavenumber satisfying $\omega'(\tilde{l})=\omega'(k-\tilde{l})$,
and $\Delta''\omega(k,l):=\omega''(l)+\omega''(k-l)\neq0$,
where ${}'$ denotes differentiation.
Eq.(\ref{eqn:a1(k)}) was obtained by using the method of stationary phase and is valid for $t\to\infty$.
This expression for $a_1$ implies that at a double resonance point $k=\bar{k}$, 
where $\Delta\omega(\bar{k},\tilde{l}(\bar{k}))=0$, $a_1$ grows as $a_1\approx O(\sqrt{t})$,
and the width of the peak around $\bar{k}$ narrows with time $t$.

\begin{figure}[!h]
%\hspace{10mm}
\begin{minipage}{0.475\linewidth}
\begin{center}
    \includegraphics[width=0.9\linewidth]
    {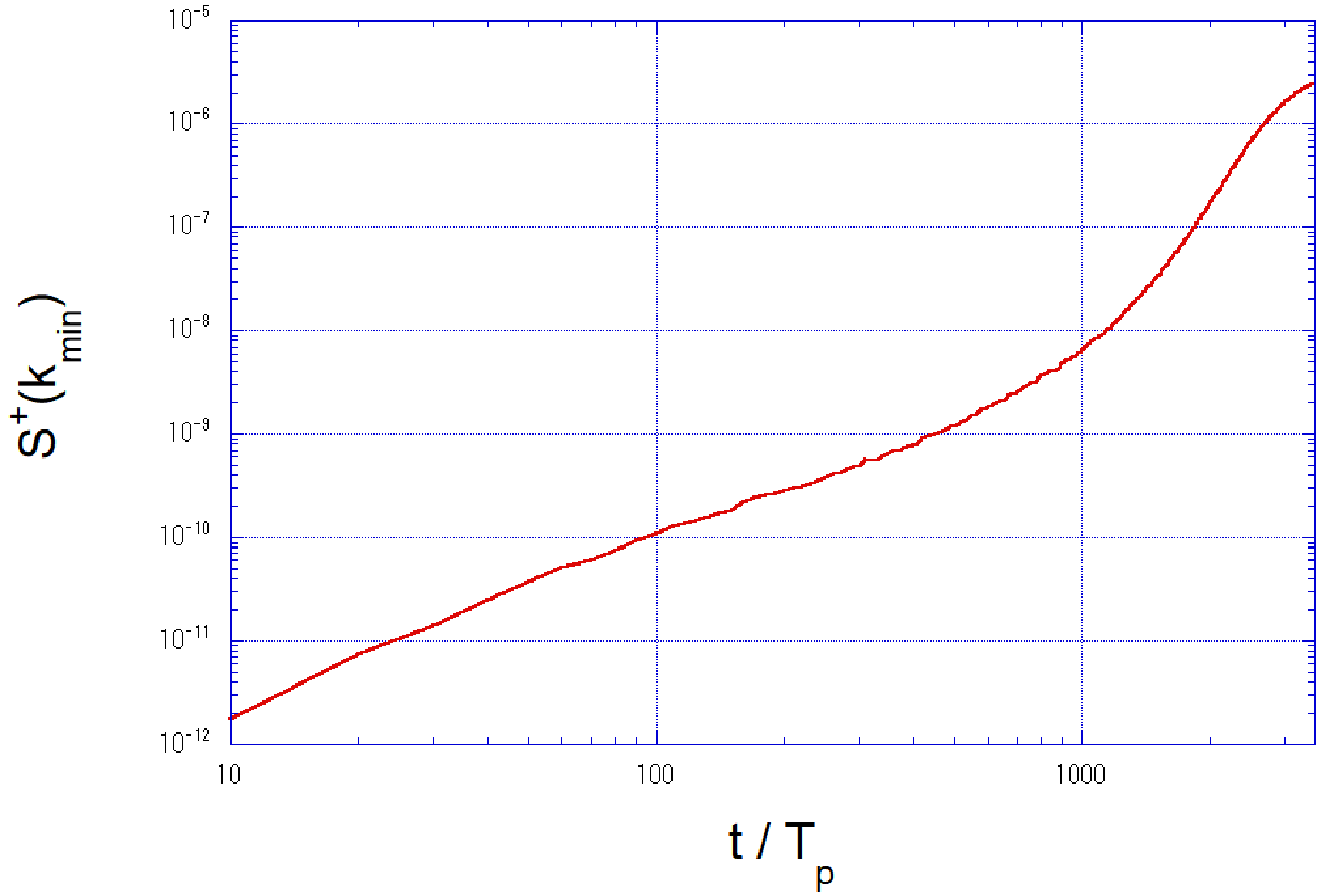}
\end{center}
\caption{Time evolution of $S^{+}(k)$ at $k_{\rm min}=1.73$ obtained by DNS of our simplified model with $E=2.5\times10^{-6}$.}
\label{fig:SS(kmin_DNS_E=2.5d-6_SS(kmin)=0)}
\end{minipage}
\hspace{0.05\linewidth}
\begin{minipage}{0.475\linewidth}
\begin{center}
    \includegraphics[width=0.9\linewidth]
    {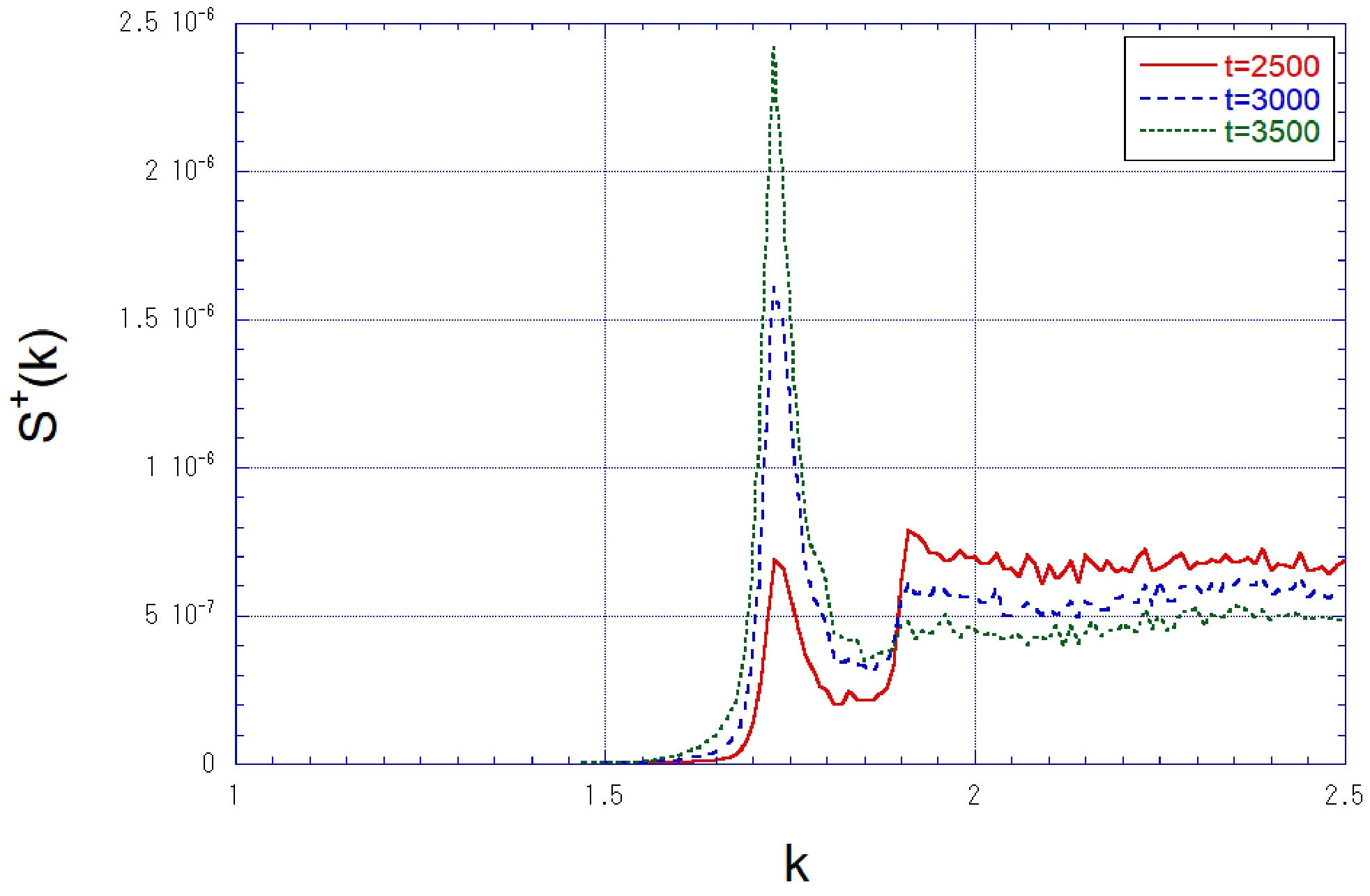}
\end{center}
\caption{Time evolution of $S^{+}(k)$ near $k_{\min}$ obtained by the same DNS as used in
Fig.\ref{fig:SS(kmin_DNS_E=2.5d-6_SS(kmin)=0)}.}
\label{fig:SpectS_DNS_E=2.5d-6_SS(kmin)=0}
\end{minipage}
\end{figure}

In the following we examine whether our numerical results are consistent with this theory.
Figure \ref{fig:SS(kmin_DNS_E=2.5d-6_SS(kmin)=0)} shows the temporal evolution of $S^{+}(k)$ at the double resonance point
$k=k_{\rm min}=1.73$ obtained from DNS of our simplified model with $E=2.5\times10^{-6}$.
In this DNS, we slightly modified the initial spectrum (\ref{eq:initial spectrum}) such that $S^{+}(k,0)=0$ for $k<1.90$.
This removes the contribution of the lowest order part $a_0$ at $k_{\rm min}$, thereby facilitating comparison
with the theoretical predictions.
The theory of \cite{Amundsen} and \cite{Amundsen-Benney} predicts that $a(k_{\rm min})$ grows as $\sqrt{t}$,
which appears to suggest that $S(k_{\rm min})$ grows in proportion to $t$.  
However, within the time range examined here, no such behavior has been observed.

%===============================================================================================
%===============================================================================================

Figure \ref{fig:SpectS_DNS_E=2.5d-6_SS(kmin)=0} shows the evolution of $S^{+}(k)$ obtained from the same DNS
near the double resonance point $k_{\rm min}$.
Judging from the fact that the growth of the peak at the double resonance point has already begun to saturate
toward the end of the computation as seen in Fig.\ref{fig:SS(kmin_DNS_E=2.5d-6_SS(kmin)=0)},
the duration of time can be considered sufficiently long.
Nevertheless, in Fig.\ref{fig:SpectS_DNS_E=2.5d-6_SS(kmin)=0}, the narrowing trend of the sharp peak
in $S^{+}(k)$ predicted by the theory is not observed.

Thus, we could not reproduce the characteristics of the evolution of modes around the double resonance point
theoretically predicted by \cite{Amundsen} and \cite{Amundsen-Benney}.
Whilst our discussion concerns the statistical quantities such as the energy spectra,
the theory in \cite{Amundsen} and \cite{Amundsen-Benney} pertains to the deterministic variables such as the
complex amplitudes of modes.
Consequently, when directly comparing the two, a more careful examination is likely to be necessary.

%=====================================================================
\section{Summary and Conclusions}
%=====================================================================
The results of this study and the conclusions derived from them can be summarized as follows.
\begin{itemize}
\item
In our previous work \cite{Tanaka-Wakayama} where we investigated numerically the energy transfer from
surface wave modes to interfacial wave modes in a two-layer fluid system, we observed an emergence and
fast growth of a sharp peak in the energy spectrum of the surface wave mode which has been brought out
by a double resonance.
This phenomenon cannot be reproduced by the conventional kinetic equation given by the wave turbulence theory (WTT).
The objective of this study is to verify whether the \lq\lq generalised" kinetic equation (GKE)
is applicable to this phenomenon, in which double resonance plays an essential role.

\item 
In order to facilitate the study, we constructed a highly simplified model of a two-layer fluid system
by extracting only specific type of three-wave interactions based on the Hamiltonian formulation of
the original system and also by replacing the complex coupling coefficient with a simple one like $V_{123}=\sqrt{|k_3|}$.

\item 
It has been confirmed by the direct numerical simulations (DNS) of this simplified model that the model can
reproduce the emergence and rapid growth of a sharp peak in the surface wave spectrum $S^{+}(k)$ at the double resonance
point $k_{\rm min}$ which is qualitatively very close to that observed in the original two-layer fluid system.
This ensures that the model can be an appropriate tool for the objective of the study.

\item 
We derived the GKE for the simplified model, performed numerical simulations based on it,
and compared the results with those from DNS of the same simplified model.
The comparison showed that the GKE can also reproduce the emergence and the fast growth of a sharp peak in $S^{+}(k)$ 
at the double resonance point $k_{\rm min}$, at least qualitatively.
However, it has been demonstrated that the GKE tends to overestimate the growth of the peak,
leading to significant discrepancies compared to the DNS results (see, for example, Fig.\ref{fig:comparison_DNS_GKE_SpectS}).
Considering that the conventional kinetic equations (\ref{eqn:kinetic equations}) predict an infinite value of $dS^{+}(k)/dt$
at $k_{\rm min}$, this may be regarded as a natural consequence.
The excessive growth at $k_{\rm min}$ also leads to deviation of the results of GKE from those of DNS
in other wavenumber regions through ordinary (i.e., single) resonant interactions.

\item
Examining the derivation process of the GKE, it is certain that the reason why GKE deviates from DNS (i.e., the correct behavior)
lies in the fact that the fourth-order cumulant has been ignored in the derivation of the evolution equation (\ref{eqn:dJ/dt})
for the third-order moment $J_{012}$ based on the assumption of weak non-Gaussianity.
In this sense, it may be worthwhile to investigate in greater detail the behaviour of $J_{012}$ as given by GKE
in future studies.

\item
Except for the vicinity of the double resonance point $k_{\rm min}$ as well as several narrow wavenumber regions
directly associated to it via ordinary resonances, the GKE results showed very good agreement with those of DNS.
In such regions where GKE agrees well with DNS,
it was also confirmed that the time scale of spectral evolution retains the similarity with respect to
the slow time scale $1/\omega_0 \epsilon^2$ as that predicted by the conventional WTT extremely well.
This might suggest that the frequency mismatch $\Delta\omega$ permitted for non-resonant interactions 
playing significant roles in the spectral evolution would scale as $O(\epsilon^2)$.
Further research into the magnitude of the permissible frequency mismatch $\Delta\omega$ would be worthwhile.

\item
Unfortunately, we were unable to confirm the theoretical predictions of \cite{Amundsen} and \cite{Amundsen-Benney} regarding
the characteristics of the growth of the modes at the double resonance point.
Since the primary objective of the present study was to investigate the applicability of the GKE to systems
exhibiting double resonance, there may have been some aspects in our method of study which are not suitable for
the purpose of comparative analysis with \cite{Amundsen} and \cite{Amundsen-Benney}.
This point warrants further consideration.
\end{itemize}

\section*{Acknowledgement}
The author wishes to express his sincere gratitude to Victor Shrira of Keele University
for drawing his attention to the fact that the appearance of a sharp spectral peak he observed in the DNS of two-layer fluid
system is related to the phenomenon of \lq\lq double resonance", and for kindly providing information on relevant literature.
He also wishes to thank Naoto Yokoyama of Tokyo Denki University for his valuable discussions.

%====================
\if0
%===============================================================================================
\subsection*{Data Availability}
The data that support the findings of this study are available from the corresponding author upon reasonable request.

\section*{Declarations}
\subsection*{Conflict of interest}
The author has no conflict of interest to declare that are relevant to the content of this article.
%===============================================================================================
\fi
%====================

%===============================================================================================

%\tableofcontents
\end{document}